\documentclass[pre, twocolumn, aps, longbibliography] {revtex4-1}

\usepackage{graphicx,epsfig}
\usepackage{amssymb, amsmath}
\usepackage{amsfonts}
\usepackage{color}
\usepackage{upgreek,textgreek}

\newcommand \beq {\begin{equation}}
\newcommand \eeq {\end{equation}}
\newcommand \be {\begin{equation}}
\newcommand \ee {\end{equation}}
\newcommand \la {\langle}
\newcommand \ra {\rangle}
\newcommand \beqn {\begin{eqnarray}}
\newcommand \eeqn {\end{eqnarray}}

\begin{document}

%%%%%%%%%%%%%%%%%%%%%%%%%%%%%%%%%%%%%%

\title{Equilibrium and dynamics of a three-state opinion model}
\author{Irene Ferri}
\author{Albert D\'iaz-Guilera}
\author{Matteo Palassini}
\email{palassini@ub.edu}

\affiliation{Departament de F\'isica de la Mat\`eria Condensada and Institute of Complex Systems (UBICS)\\ 
Universitat de Barcelona, 08028 Barcelona, Spain}

\date{\today}

\begin{abstract}
{\bf Abstract:} We introduce a three-state model to study the effects of a neutral party on opinion spreading, in which the tendency of agents to agree with their neighbors can be tuned to favor either the neutral party or two oppositely
polarized parties, and can be disrupted by social agitation mimicked as temperature. We study the equilibrium phase 
diagram and the non-equilibrium stochastic dynamics of the model with various analytical approaches and with
Monte Carlo simulations on different substrates: the fully-connected (FC) graph, the one-dimensional (1D) chain, 
and Erd\"os-R\'enyi (ER) random graphs. 
We show that, in the mean-field approximation, the phase boundary between the disordered and polarized phases is characterized by a tricritical point. On the FC graph, in the absence of social agitation, kinetic barriers prevent the system from reaching optimal consensus. 
On the 1D chain, the main result is that the dynamics is governed by the growth of opinion clusters.
Finally, for the ER ensemble a phase transition analogous to that of the FC graph takes place, but
now the system is able to reach optimal consensus at low temperatures, except when the average connectivity
is low, in which case dynamical traps arise from local frozen configurations.

\end{abstract}

\maketitle

%%%%%%%%%%%%%%%%%%%%%%%%%%%%%%%%%%%%%%%%%%%%%%%%%%%%%%%%%%%%%%%%%%%%%%%%%%%
\section{Introduction}
Within the field of complex systems, social questions are perhaps the most elusive, as the agents involved (humans) exhibit a sophisticated individual behavior, not easily reducible
to a few analyzable parameters. Nevertheless, many models have been proposed to capture different aspects of societal interaction, such as bipartidism \cite{Abelson64, Deffuant2002, Banisch2018, Yang_2020}, gerrymandering \cite{jacobs2018partial, Stewart2019, Bergstrom2019}, or echo chambers formation \cite{Gaisbauer_2020, Kureh_2020, Baumann2020}.

The consensus problem on a given social question, such as which kind of energy is the most suitable for subsistence or which political party should govern, has been addressed using a variety of agent-based models, both discrete, such as the voter \cite{Holley1975, Suchecki2005, Kureh_2020}, Ising \cite{Li2019} or Potts models \cite{Schulze_2005}, and continuous, such as the Deffuant model \cite{Deffuant2000}. Continuous models often predict the formation of opinion clusters \cite{Weisbuch2004, Sobkowicz2015}, thereby reinforcing a discrete description of the opinion space. 

A common goal in many of these works is to understand the transition between an initial disordered 
state, in which opinions are random and uncorrelated, to a state in which agents exhibit some kind of local or global consensus.
The simplest case occurs when there are only two opinions, 
as in polarized situations with a clear bipartidist scenario. In other situations,
however, 
it is more realistic to consider at least one intermediate or  neutral state representing, for example, centrists or undecided voters. Several three-state models have been proposed,
using various approaches for introducing the neutral state.
Some of these models 
prevent agents that hold extreme opinions from interacting directly, forcing them to pass through the neutral opinion.
For instance, Vazquez and Redner \cite{Vazquez2004} study a stochastic kinetic model in the mean-field approximation, and find that the final configuration depends strongly on the initial proportion of agents in each state. Along similar lines, the authors of Ref.\cite{Svenkeson2015} incorporate temperature and find a phase transition analogous to that of the Ising model. 

In this paper we propose a three-state Hamiltonian agent-based model for opinion spreading 
in which agents interact in a pairwise manner that tends to promote consensus,
with a tunable neutrality parameter that controls the relative preference for the neutral state over the polarized states. Agents can change their opinion according to a stochastic dynamics in which the effects of social agitation are taken into account by mimicking them as a temperature.

The model can be mapped to a special case of the
Blume-Emery-Griffiths (BEG) model \cite{Blume1971} from condensed matter physics.
Other variants of the BEG model have
been applied to sociophysics before \cite{Yang2010, Fernandez2016}, but our model 
allows to study directly the role of the neutral state in the dynamics of the opinion
formation. 

The geometry of social structures is crucial in opinion formation and 
 other contemporary questions such as pandemic spreading,
 economics \cite{Souma_2003, Li_2003, DaSilva_2022} and smart cities design \cite{Lu_2015, AlSonosy_2018}.
In order to understand the role
of the network geometry in the dynamics of opinion formation, we embed our 
model on different types of networks:  the fully connected (FC) graph, the one-dimensional (1D) chain,
and Erd\"os-R\'enyi (ER) random graphs.
We study with different analytical approaches the equilibrium phase diagram of the model, and use Monte Carlo (MC) simulations at zero and non-zero temperature 
to investigate the stochastic evolution of the population
starting from random configurations.

The paper is organized as follows. In Section \ref{The Model} we introduce the model, 
placing it in a social context and discussing its general features. In Section \ref{MF} we determine the equilibrium phase diagram in the mean-field approximation, which is characterized by a phase boundary with a tricritical point.
In Section \ref{FCG} we analyze the 
zero-temperature dynamics on the FC graph and identify the basins of attraction
of the absorbing states. We predict, and confirm via MC simulations, that random configurations always evolve towards an all-neutral state, and a finite temperature is necessary to overcome the barriers towards equilibrium. 
In Section \ref{One} we derive an exact solution for the 1D chain, and obtain
the scaling of the magnetization from a domain-growth argument, which is validated by our MC results.
In Section \ref{Erdos}, devoted to ER graphs, we determine the transition temperature using the annealed mean-field approximation, and show that for low connectivities the system gets stuck in dynamical traps, while for high connectivities it is always able to reach the optimal configuration. Finally, in Section \ref{conclusions} we present our conclusions. The Appendices contain details of the analytical calculations and numerical methods.

%%%%%%%%%%%%%%%%%%%%%%%%%%%%%%%%%%%%%%%%%%%%%%%%%%%%%%%%%%%%%%%%%%%%%%%%%%%
%%%%%%%%%%%%%%%%%%%%%%%%%%%%%%%%%%%%%%%%%%%%%%%%%%%%%%%%%%%%%%%%%%%%%%%%%%%
\section{The Model} \label{The Model}
\begin{figure}[h!]
\centering
\includegraphics[width=\columnwidth]{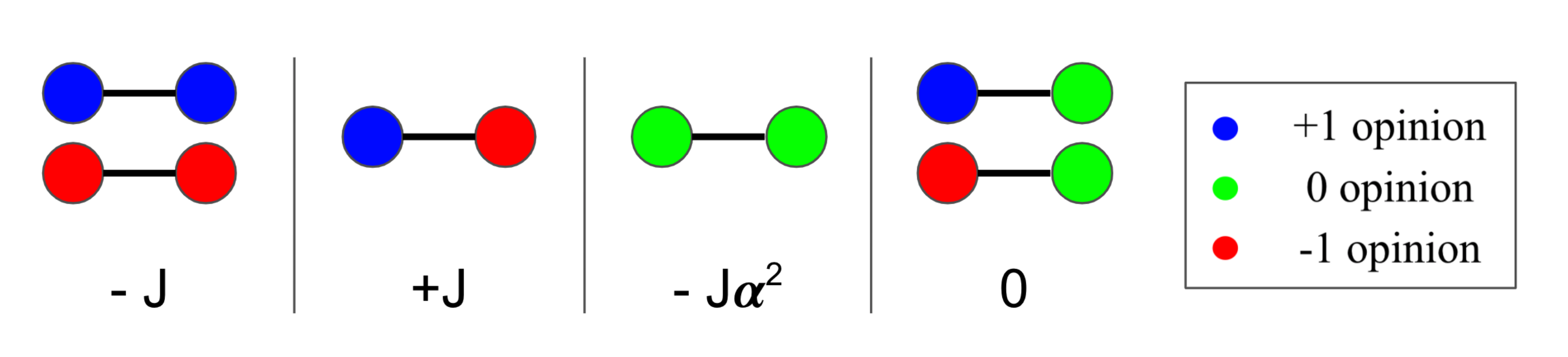}
\caption{\small{Contribution of a pair of interacting agents to the energy of the system.}}
\label{contributions}
\end{figure}

We propose an agent-based model with a discrete opinion space. Agents live on the $N$ nodes of an undirected graph,
and their opinions are represented by two-dimensional vectors ${\bf S}_i$ ($i=1,\dots,N$) that can take three
orientations:

\begin{itemize}
\item ${\bf S}_i  =  (1, 0)$;  positive opinion / rightist
\item ${\bf S}_i  =  (0, \alpha)$;  neutral opinion / centrist
\item ${\bf S}_i   = (-1, 0)$;   negative opinion / leftist
\end{itemize}
where $\alpha$ is the neutrality parameter. 
We assume that the agents prefer to agree with their neighbors so as to minimize,
in the absence of social agitation, the
following cost function, or Hamiltonian:
\begin{equation}
H =-J \displaystyle\sum_{\langle i , j\rangle} {\bf S}_i \cdot {\bf S}_j,
\label{eq:hamiltonian}
\end{equation}
where $J>0$ and the sum runs over all the undirected edges of the graph. In the 
optimal configuration, i.e. the
ground state of the Hamiltonian, agents reach consensus on the neutral opinion
if $\alpha>1$, or on one of the two polarized opinions if $\alpha<1$.
The possible values of the interaction energy between two agents,
$-J {\bf S}_i \cdot {\bf S}_j$ are shown in Fig.~\ref{contributions}. The
quantity $J (\alpha^2 -1)$ thus measures the reward of the neutral opinion over
the polarized ones.
We assume that the system is in contact with a thermal bath at temperature $T$,
which should be understood as a coarse-graining of all the sources of noise
(e.g. social agitation) that affect individual opinions.

An equivalent way to express the above Hamiltonian  is to
replace the state vectors  with scalar variables
taking the values $\sigma_i \in \{1, 0, -1\}$, which gives
\begin{equation}
H=-J\displaystyle\sum_{\langle i , j \rangle} \left[\sigma_i \sigma_j + \alpha^2 (\sigma_{i}^{2} - 1)(\sigma_{j}^{2} - 1)\right] \,.
\label{eq:hamiltonian_2}
\end{equation}

In this form, the model can be seen as a special case of 
the BEG model \cite{Blume1971}, which in its general form is described by the Hamiltonian 
\begin{equation}
H^{BEG} = -J \displaystyle \sum_{\langle i , j \rangle} \sigma_{i} \sigma_{j} - K \displaystyle\sum_{\langle i , j\rangle} \sigma_{i}^{2} \sigma_{j}^{2} + \displaystyle  \sum_{i} \Delta_i \sigma_{i}^{2} + C \,.
\label{eq:BEG}
\end{equation}
In fact, we recover Eq.\eqref{eq:hamiltonian_2} by setting
\begin{equation}
K = J\alpha^{2} , \quad \Delta_i = k_i K, \quad C = -N z K \mathbin{/} 2,
\label{eq:mapping}
\end{equation}
where $k_i$ is degree of node $i$, i.e. the number of nodes connected to it, 
and $z=\sum_i k_i/N$ is the average degree of the graph.
Note that the standard BEG model has a unique value $\Delta_i = \Delta$ for all $i$.
Therefore, for graphs with constant degree $k_i=z$, 
we can obtain the phase diagram of our model 
by projecting that of the BEG model, which is defined in the 
three-dimensional parameter space 
($K/J$, $\Delta/J$, $T/J$),
onto the two-dimensional semiplane ($\alpha, T/J$) defined by
Eq.\eqref{eq:mapping}.

We assume that the system evolves stochastically via either one of two
common discrete-time Markov processes, the Metropolis and the Glauber dynamics,
both described in Appendix A.  
For $T>0$, both dynamics converge, given sufficient time, to a stationary state in which the probability of a configuration 
$\sigma=\{\sigma_i\}_{i=1}^N$ is given by the Boltzmann distribution
$p(\sigma)\propto \exp[- H(\sigma)/T]$.
At $T=0$, depending on the graph and initial conditions, they can either converge to the 
ground state of the Hamiltonian, or get trapped forever in metastable configurations, as we will discuss later.
We perform MC simulations with both dynamics starting,
unless otherwise specified, from a random configuration in which the
agents independently take one of the
three opinion states with uniform probability (i.e. sampling from the infinite-temperature Boltzmann distribution),
and then quenching the system instantaneously to the desired temperature
$T$ and letting it evolve. 

The order parameters of the model are the Ising-like magnetization $m = \sum_i \la \sigma_i\ra / N$, 
namely the difference between the fractions of rightists and leftists, and 
the fraction of neutral agents, $n_0 =1-\sum_i \la \sigma_i^2\ra / N$.  
Here $\la \dots\ra$ denotes the expectation with respect to the Boltzmann distribution.

In the rest of the paper we will use dimensionless units for the energy and temperature such that $J=1$.
We can anticipate some general features of the equilibrium phase diagram  in the plane $(\alpha,T)$. 
For $\alpha < 1$, the ground state is ferromagnetic, namely the agents achieve a global
consensus either in the positive or in the negative state ($m=\pm 1, n_0=0$).
For $\alpha>1$,  all agents assume the neutral opinion ($m=0, n_0=1$) in the ground state.
Hence, moving along the zero-temperature axis we encounter a discontinuous phase transition at $\alpha=1$.

For $\alpha=0$, the model can be thought of as an Ising model with vacancies. 
Hence, in the thermodynamic limit it will
generically display a continuous phase transition 
in the Ising universality class
at a critical temperature $T_c^0 > 0$, between a low-temperature polarized (ferromagnetic) phase,
characterized by $m \neq 0$,  and a high-temperature 
disordered (paramagnetic) phase, in which $m = 0$.

We thus generally expect a phase boundary in the plane $(\alpha,T)$ connecting the two points
$(\alpha=0, T=T_c^0)$ and  $(\alpha=1,T=0)$.
Since the transition is discontinuous at one end of the boundary and continuous at the other,
we also expect a tricritical point $(\alpha_{tc},T_{tc})$ at some point along
the boundary, separating a line of continuous transitions at $T=T_c(\alpha)$ for  $0\leq \alpha \leq \alpha_{tc}$ from
a line of discontinuous transitions at $T=T_d(\alpha)$ for $\alpha_{tc}< \alpha \leq 1$. 
Such a phase boundary was indeed observed for a different projection of the BEG model, the
Blume-Capel model, on various types of random graphs, including some cases in which
the phase boundary is reentrant \cite{Leone}.
Exceptions to this scenario are represented by 1D systems with short-range interaction, in which $T_c^0=0$, since no long-range
order can survive at finite temperature, and by 
graphs with a degree distribution falling more slowly than 
$k^{-3}$ for large $k$, in which the system is known to remain
ferromagnetic at all temperatures for $\alpha =0$ \cite{Leone} (we expect this to remain true for all $\alpha < 1$).

%%%%%%%%%%%%%%%%%%%%%%%%%%%%%%%%%%%%%%%%%%%%%%%%%%%%%%%%%%%%%%%%%%%%%%%%%%%
%%%%%%%%%%%%%%%%%%%%%%%%%%%%%%%%%%%%%%%%%%%%%%%%%%%%%%%%%%%%%%%%%%%%%%%%%%%
\section{Mean-field phase diagram} \label{MF}
\begin{figure}[h!]
\centering
\includegraphics[width=\columnwidth]{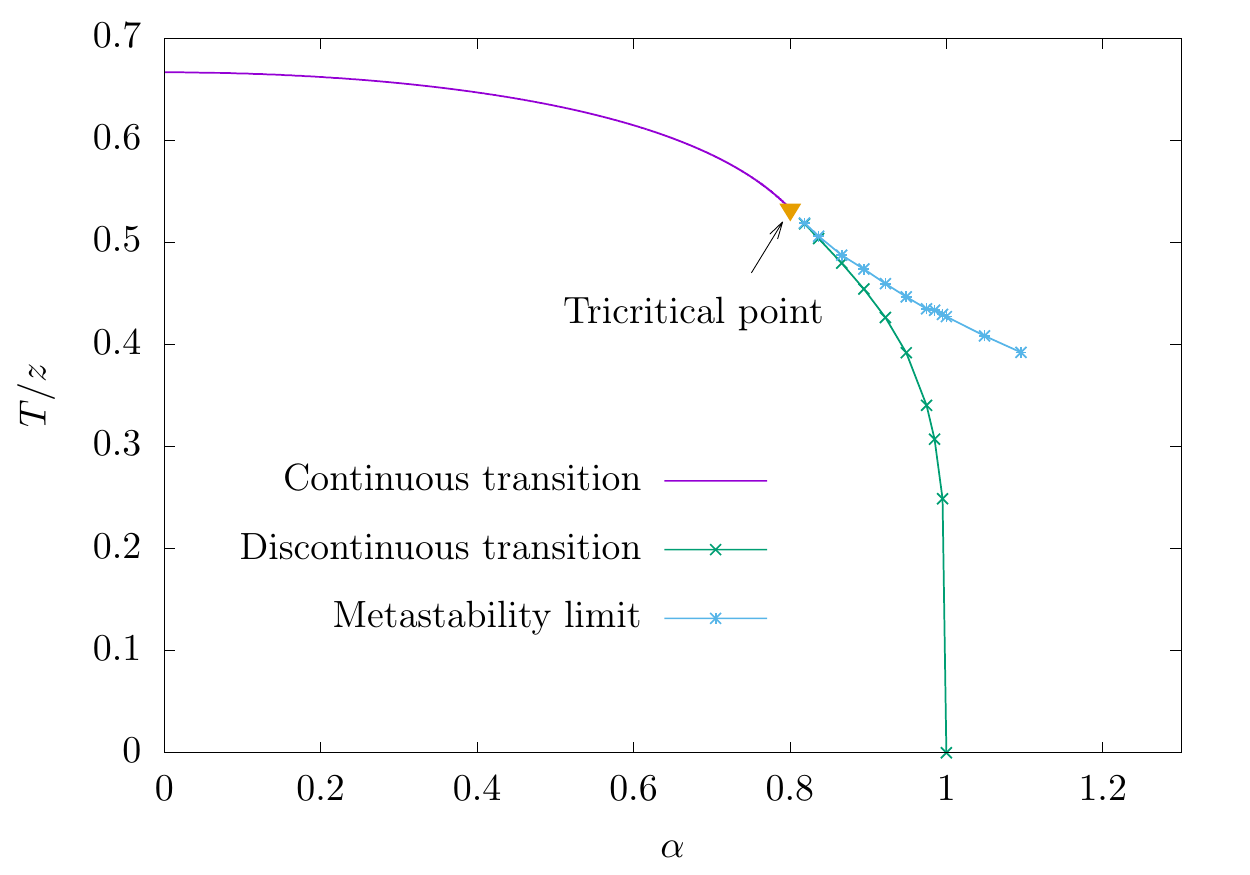}
\caption{\small{Mean-field phase diagram. The line of continuous 
transitions $T = \beta_c(\alpha)^{-1}$ for $\alpha< \alpha_{tc}$ 
is given by Eq.\eqref{eq:boundary}. For $\alpha>\alpha_{tc}$, the points 
are obtained numerically as explained in the text, the lines being only a guide to the eye.}}
\label{fully_phase}
\end{figure}

The mean-field approximation
provides a useful first understanding of the equilibrium behavior of the model.
In this approximation, the free-energy per spin is given by the minimum
of a free-energy function ${\cal L}(m,n,\beta)$ with respect to $m$ and $n = 1 -n_0$,
where $\beta = 1/T$ is the inverse temperature.
As shown in Appendix A, 
the
stationarity conditions ${\partial {\cal L}}/{\partial m}= 
{\partial {\cal L}}/{\partial n}=0$  
give two coupled self-consistent equations (SCEs),
\begin{eqnarray}
m &=& \frac{2e^{\beta z \alpha^2(n-1)} \sinh(\beta z m)}{1 + 2e^{\beta z \alpha^2(n-1)} \cosh(\beta z m)},
\label{eq:SCE_m}  \\
n &=&  \frac{2e^{\beta z \alpha^2(n-1)} \cosh(\beta z m)}{1 + 2e^{\beta z \alpha^2(n-1)} \cosh(\beta z m)}.
\label{eq:SCE_n}
\end{eqnarray}
By expanding them for small $m$, 
we find a line of continous transitions 
between the disordered phase
and the polarized phase, at an inverse critical temperature $\beta_c(\alpha)$ given by
\begin{equation}
\alpha^2 = \frac{1}{\beta_c(\alpha) z - 1} \ln \left[2(\beta_c(\alpha) z - 1)\right]\,.
\label{eq:boundary}
\end{equation}
In particular, we have $\beta_c(0)^{-1}=2 z /3$,
which is below the critical temperature 
$T_c = z$ of the Ising model in the mean-field approximation. This is
because, even if at $\alpha=0$ the neutral state does not contribute to the energy,
it brings an additional entropy that
destabilizes the polarized phase.
The line of critical points, shown in Fig.~\ref{fully_phase}, 
ends at a tricritical point
$(\alpha_{tc},\beta_{tc})$ determined by the condition
\begin{equation}
2 \ln [2(\beta_{tc} z - 1)] = 3 - \beta_{tc} z 
\label{eq:fully_tricritical}
\end{equation}
By solving this numerically 
we obtain $\beta_{tc}^{-1} = 0.532573 \, z $, and substituting this value into Eq.\eqref{eq:boundary}
gives $\alpha_{tc} = 0.800354$.

For $\alpha>\alpha_{tc}$, the phase diagram displays a line of discontinuous
transitions at inverse temperature $\beta_d(\alpha)$, which we locate
by finding numerically the values of $m$, $n$, $\beta$ that satisfy simultaneously Eqs.\eqref{eq:SCE_m} and \eqref{eq:SCE_n}, together with the condition of equality between 
the free energies of the ferromagnetic and paramagnetic phases.
The resulting phase boundary is shown in Fig.~\ref{fully_phase}. 
Also shown in the figure is the limit of metastability of the ferromagnetic phase, 
namely the value of $T$ above which the SCEs no longer admit a solution with $m\neq 0$.

\begin{figure}[h!]
\centering
\includegraphics[width=\columnwidth]{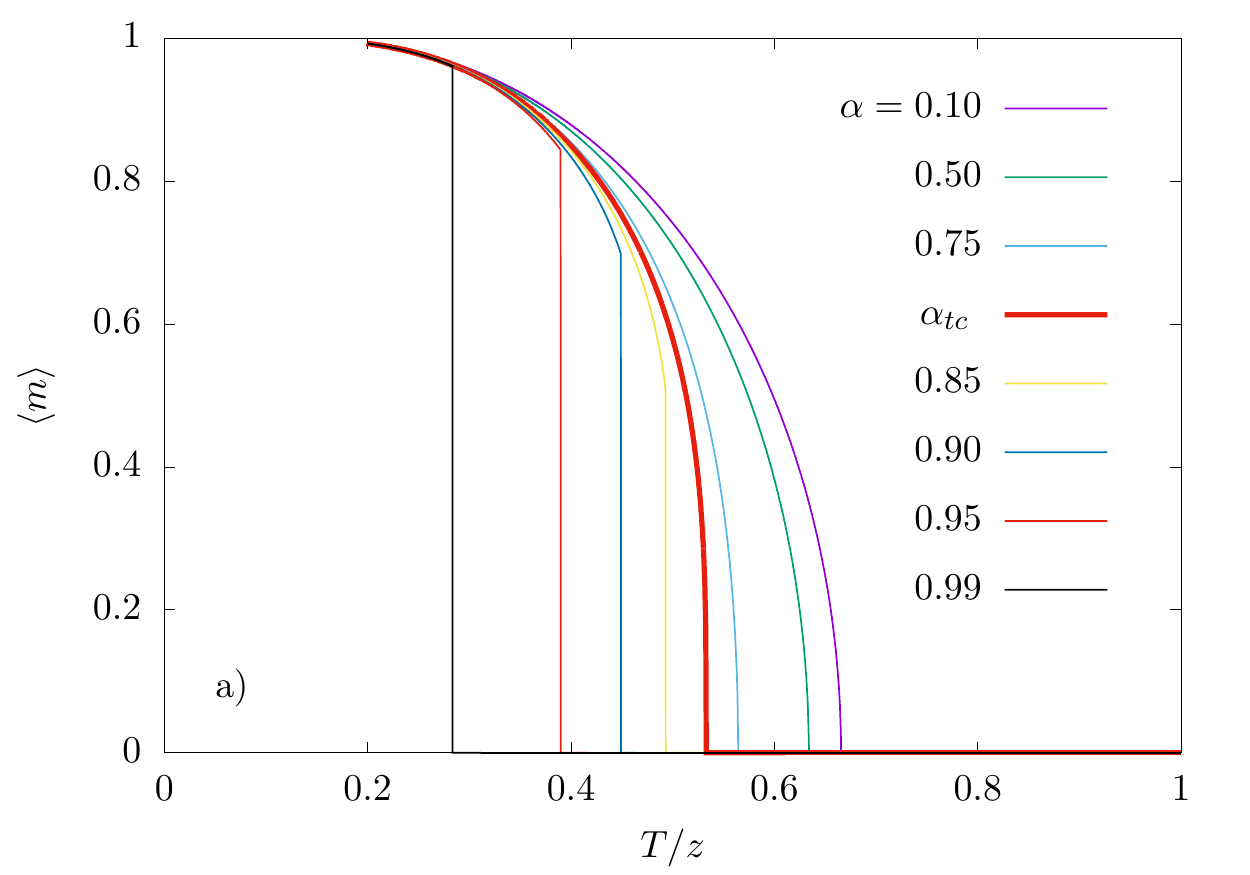}
\includegraphics[width=\columnwidth]{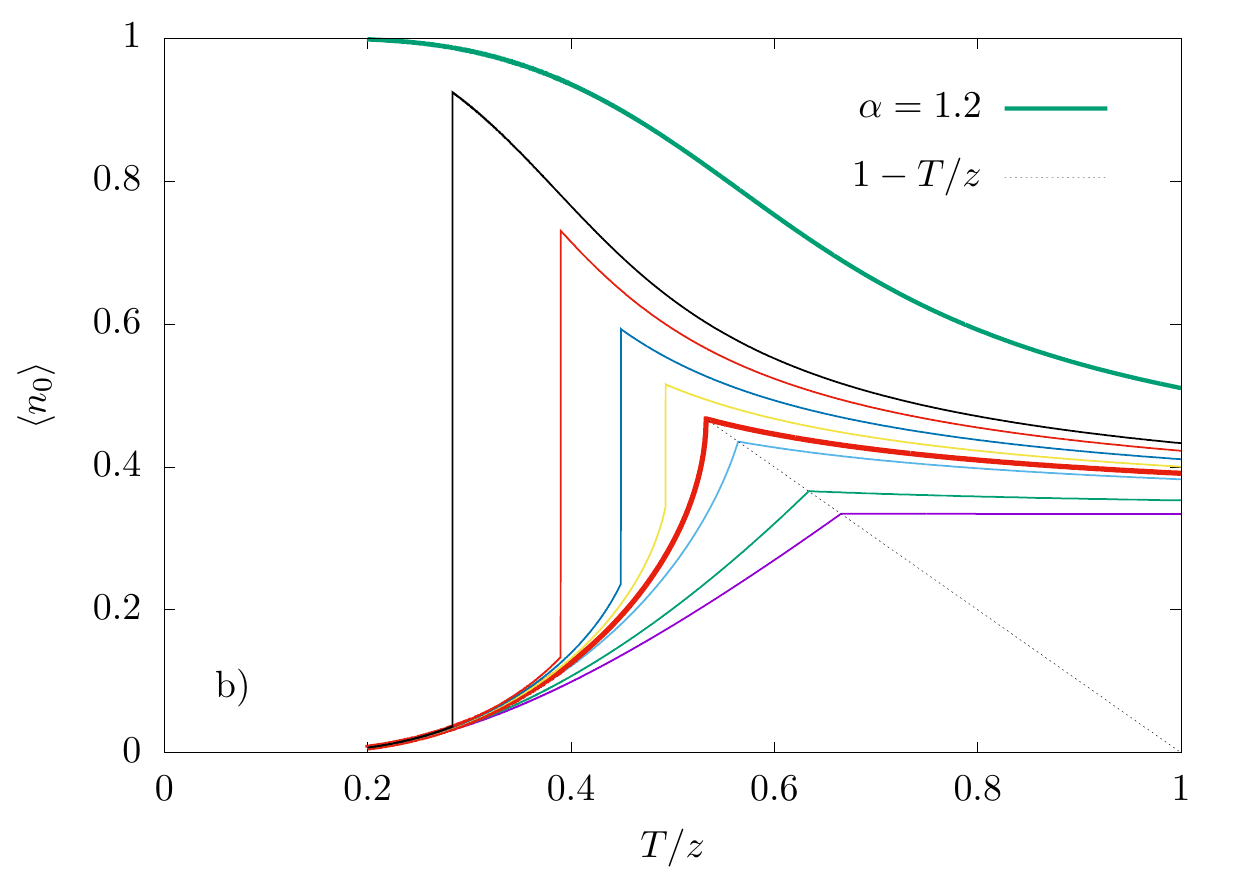}
\caption{\small{Mean-field results for the equilibrium values of the magnetization (a) and
the fraction of neutral sites (b), as a function of the temperature and for different values of $\alpha$. 
The thick red line corresponds to $\alpha=\alpha_{tc}$, separating the continuous
transition from the discontinuous transition.
In b) the thick green curve corresponds to $\alpha=1.2$, and the other curves are as in a).}}
\label{fig:MF}
\end{figure}

The equilibrium values $\la m\ra$ and $\la n_0\ra = 1-\la n\ra$, obtained by solving numerically
Eqs.\eqref{eq:SCE_m} and \eqref{eq:SCE_n},
are shown in Fig.~\ref{fig:MF} as a function of temperature.
For $\alpha < \alpha_{tc}$, the magnetization vanishes continuously at the critical temperature, and is described by the usual mean-field critical and tricritical exponents as $T\to T_c(\alpha)^-$, namely
$\langle m\rangle \sim (T_c(\alpha) - T)^{1/2}$ for $\alpha<\alpha_{tc}$
and $\langle m \rangle \sim (T_{tc} - T)^{1/4}$ for $\alpha=\alpha_{tc}$. The
fraction of neutral agents
$\la n_0\ra$ increases with $T$ up to the critical point $T_c(\alpha)$, then it decreases monotonically
towards $\langle n_0\rangle =1/3$ in the $T\to \infty$ limit, in which the three states become equiprobable. We note that at $T=T_c(\alpha)$ we have $\langle n_0\rangle = 1 - T_c(\alpha)/z$.

For $\alpha_{tc}<\alpha<1$, $\la m\ra$ decreases monotonically with $T$ and jumps to zero at $T=T_d(\alpha)$, while $\langle n_0\rangle$ has a strongly non-monotonic temperature dependence: starting from
$\langle n_0 \rangle=0$ at  $T=0$, it increases slowly with $T$, then it jumps to a large value at the discontinuous
transition, before decreasing monotonically towards $1/3$. 

Finally, for $\alpha \geq 1$, we have $\langle m\rangle=0$ at all temperatures, and $\langle n_0\rangle$ decreases monotonically with $T$, starting from $\langle n_0\rangle=1$ at $T=0$, 
since in the ground state all agents are in the neutral state.

From a social point of view, we see that agents agree on one of the polarized opinions when coupling dominates over temperature, while at intermediate levels of upheaval above the discontinuous transition they have a neutral preference.

The mean-field approximation is exact for the FC graph in the
$N\to\infty$ limit, formally setting $z=1$ (see Section \ref{FCG}).
In addition, the qualitative features of the
phase diagram and of the behavior of $\la m\ra$ and $\la n_0\ra$ described above 
will hold in a large class of graphs.
In particular, for $d$-dimensional regular lattices 
we expect the same qualitative phase diagram when $d\geq 2$ (no finite-temperature phase transition can exist
for $d=1$), with mean-field critical exponents for $d>4$ and non-mean-field exponents of the Ising 
universality class for $d=2, 3$.

For random graphs, the mean-field approximation is not exact and various other theoretical approaches
have been developed. For the Ising model, it  was found, using the replica method,
that if the degree distribution falls off as 
$k^{-5}$ for large $k$ or faster (which includes the case of random-regular and ER graphs),
a continuous transition with mean-field exponents takes place. On the other hand,
for $\gamma\leq 5$ different scenarios depending on $\gamma$ are observed \cite{Leone}. 
In the Blume-Capel model, a tricritical point was found using
an annealed mean-field approximation (see Section \ref{sec:AMF}) \cite{DeMartino2012}.

\section{Fully-connected graph} \label{FCG}

Next, we discuss the stochastic dynamics of the model on the FC graph, in which each 
agent interacts with all other agents. In this case, in order for the Hamiltonian to be
extensive (i.e. proportional to $N$), we must replace  the 
coupling constant $J$ in Eq.\eqref{eq:hamiltonian} by $J/N$. Since
the degree is $z=N-1$, for large $N$ we have $z J/N= J$, which is equivalent to 
setting $z=1$ in the mean-field solution. The Hamiltonian can then 
be written, up to terms of order $1/N$ and recalling that $J=1$ in our units, as
\beq
H = -\frac{N}{2} (m^2 + \alpha^2 n_0^2) \,.
\eeq
The dynamics can then be represented as the evolution of a
point $(m,n_0)$ inside the triangle of vertices $(1,0),(-1,0),(0,1)$
shown in Fig.~\ref{flow}. We will refer to any such point
as the macrostate of the system, to distinguish it 
from the microscopic configuration of the $N$ agents.

\subsection{Zero-temperature dynamics}\label{sec:T0FC}

\begin{figure}[h!]
\centering
\includegraphics[width=1.02\columnwidth]{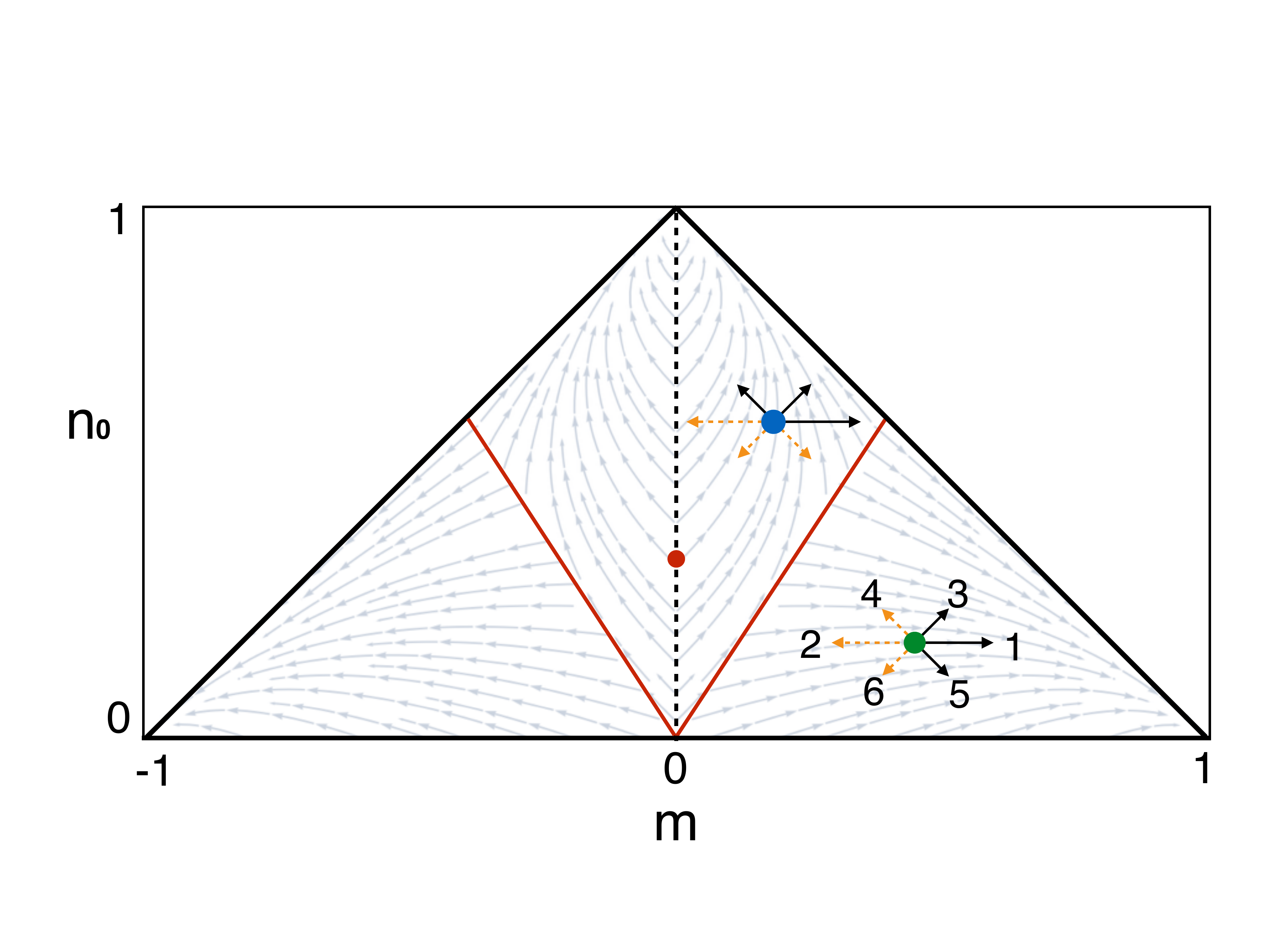}
\caption{\small{Evolution of the Metropolis dynamics on the FC graph. The red lines 
are the separatrices $m=\pm n_0 \alpha^2$ of the three basins of attraction (shown here for $\alpha=0.8$). The red point shows 
the equiprobable macrostate $m=0, n_0=1/3$. The solid black arrows 
represent the allowed moves ($\Delta H\leq 0$) from a macrostate below (green point) and above (blue point) the
separatrix for $m>0$. The dashed orange arrows represent the forbidden moves ($\Delta H >0$). The light grey lines show the average direction of the
evolution at each point.}}
\label{flow}
\end{figure}

If we sample the initial configuration of the 
$N$ agents independently and uniformly at random
among the three states,
the probability distribution of the initial macrostate is
\beq 
p(m,n_0) = 3^{-N} \frac{N!}{(N n_{+})! \, (N n_-)! (N n_0)!}, 
\eeq
where $n_{\pm} = (1-n_0\pm m)/2$ is the fraction of agents in the positive and negative state, respectively.
For large $N$, one has $p(m,n_0) \sim \exp [N f(m,n_0)]$,
where $f(m,n_0)$ is a large-deviation function that has a maximum at $m=0, n_0=1/3$, which corresponds to
equiprobable opinions, and is shown by the red point in Fig.~\ref{flow}. The probability of any  macrostate
different than the equiprobable one decreases exponentially with $N$. 

It is nevertheless interesting to analyze the fate of the system prepared in 
an arbitrary macrostate $(m,n_0)$, even if exponentially rare, as it might be relevant from a social viewpoint.
At zero temperature, with both the Glauber and the Metropolis dynamics the system can
only evolve from any given configuration by moves that do not increase $H$.
In the following we will focus on the Metropolis dynamics, in which an elementary move consists
in choosing an agent at random and proposing to change its state with probability 1/2 to 
either of the two states different from the current one,
and accepting the proposal if the energy change $\Delta H$ is negative or zero, in which
case a new macrostate $(m^\prime,n_0^\prime)= (m+\Delta m, n_0+\Delta n_0)$ is reached.

Since there are two possible proposals starting from each of the three opinion states, 
there are six possible moves, except at 
the edges of the triangle where some transitions are forbidden, and at the vertices
where they are all forbidden. The six moves are listed in Table \ref{moves}
together with their proposal probability $w_r$ ($r=1,\dots,6$), the displacement vector 
$(\Delta m^r, \Delta n_0^r)$ multiplied by $N$,
and the energy change $\Delta H_r$ (up to terms of order $1/N$) upon accepting the move $r$.
For example, move $r=1$ consists in picking at random an agent $i$ that is currently
in the state $\sigma_i=-1$ and proposing to change its state to $\sigma_i=+1$. The probability
to propose this move is $n_{-}/2$ (since the fraction of agents with $\sigma_i=-1$ is $n_{-}$
and we can choose to go to $\sigma_i=1$ or 0 with probability 1/2). The total magnetization changes by $+2$,
the number of neutral sites remains unchanged, and the energy changes by $-[(M+2)^2-M^2]/(2 N)
= - (2 m + 2/N)$.

All macrostates are unstable with respect to downhill moves (i.e. moves with $\Delta H_r < 0$) except
the vertices of the triangle, which are absorbing states.
The lines $m=\pm \alpha^2 n_0$ are the separatrices of the basins of attractions of the
absorbing states, as they discriminate between different sets of allowed moves. 

A full analysis of the master equation associated to the stochastic dynamics would allow to
determine the probability to end up in each of the absorbing states
for a given initial macrostate, but is beyond the scope of the paper. 
In Fig.~\ref{flow} we display instead the flux lines corresponding to the {\em average} 
displacement, $(\langle \Delta m \rangle, \langle \Delta n_0 \rangle)$, given by
\beq
\langle \Delta m \rangle = \sum_{r =1}^6 w_r \Delta m_r \theta(- \Delta H_r)\, ,
\eeq
and similarly for $\langle \Delta n_0 \rangle$, where
$\theta$ is a Heaviside function ensuring that only downwill moves contribute.
We can see that below the separatrices the average
flux flows to the polarized $m=\pm 1$ states, and above the separatices it flows to 
$n_0=1$.                        

In more detail, we observe that if $\alpha < 1$ and $m>0$ (the case $m<0$ follows by symmetry)
all the allowed moves from a point below the separatrix (moves 1, 3 and 5, as shown in Fig.~\ref{flow})
displace the point further away from the separatrix.
Thus all the points below the separatix belong
to the basin of attraction of the $m=1$ absorbing state and have probability zero to flow 
to $n_0=1$.
In contrast, a point above the separatrix can move away from the separatrix with probability $n_+/2$ (move 4)
or move towards it with total probability $n_-$ (the sum of the probabilities of moves 1 and 3). If $n_0 < 1-3 m$, the former probability is smaller than the latter, thus 
the preferred direction is towards the separatrix. Therefore, a starting point in the triangle
defined by $m/\alpha^2 < n_0 < 1-3 m, m>0$ will have a non-zero probability 
to end up in the ``wrong''  absorbing state $m=1$. However, since there are many more paths 
towards $n_0=1$ than towards $m=1$, this probability will in fact decrease exponentially with $N$.

\begin{table}
\begin{center}
\begin{tabular}{|c|c| c| c |c|}
\hline
$r$&Transition & $w_r$ & $ N (\Delta m^r, \Delta n_0^r)$ & $\Delta H_r$ \\ 
\hline
1&$- \longrightarrow +$&$n_-/2$ &$(2,0) $ & $ -2 m $ \\[1mm]
2&$+ \longrightarrow -$&$n_+/2$ &$(-2,0) $ & $  2 m $ \\[1mm]
3&$- \longrightarrow 0$&$n_-/2$ &$(1,1) $ & $ -m - \alpha^2 n_0 $ \\[1mm]
4&$+ \longrightarrow 0$&$n_+/2$ &$(-1,1) $ & $ m -\alpha^2 n_0 $ \\[1mm]
5&$0 \longrightarrow +$&$n_0/2$ &$(1,-1) $ & $ -m+\alpha^2 n_0 $ \\[1mm]
6&$0 \longrightarrow -$&$n_0/2$ &$(-1,-1) $ & $ m+\alpha^2 n_0 $ \\[1mm]
\hline
\end{tabular}
\end{center}
\caption{Allowed transitions for the Metropolis algorithm. The second column shows the transition between two spin values. $w_r$ is the probability of proposing the transition, $N\Delta m^r$, $N\Delta n_0^r$,
$\Delta H_r$ are the changes in total magnetization, number of neutral spins, and Hamiltonian upon executing the transition $r$.}
\label{moves}
\end{table}

\begin{figure}[h!]
\centering
\includegraphics[width=\columnwidth]{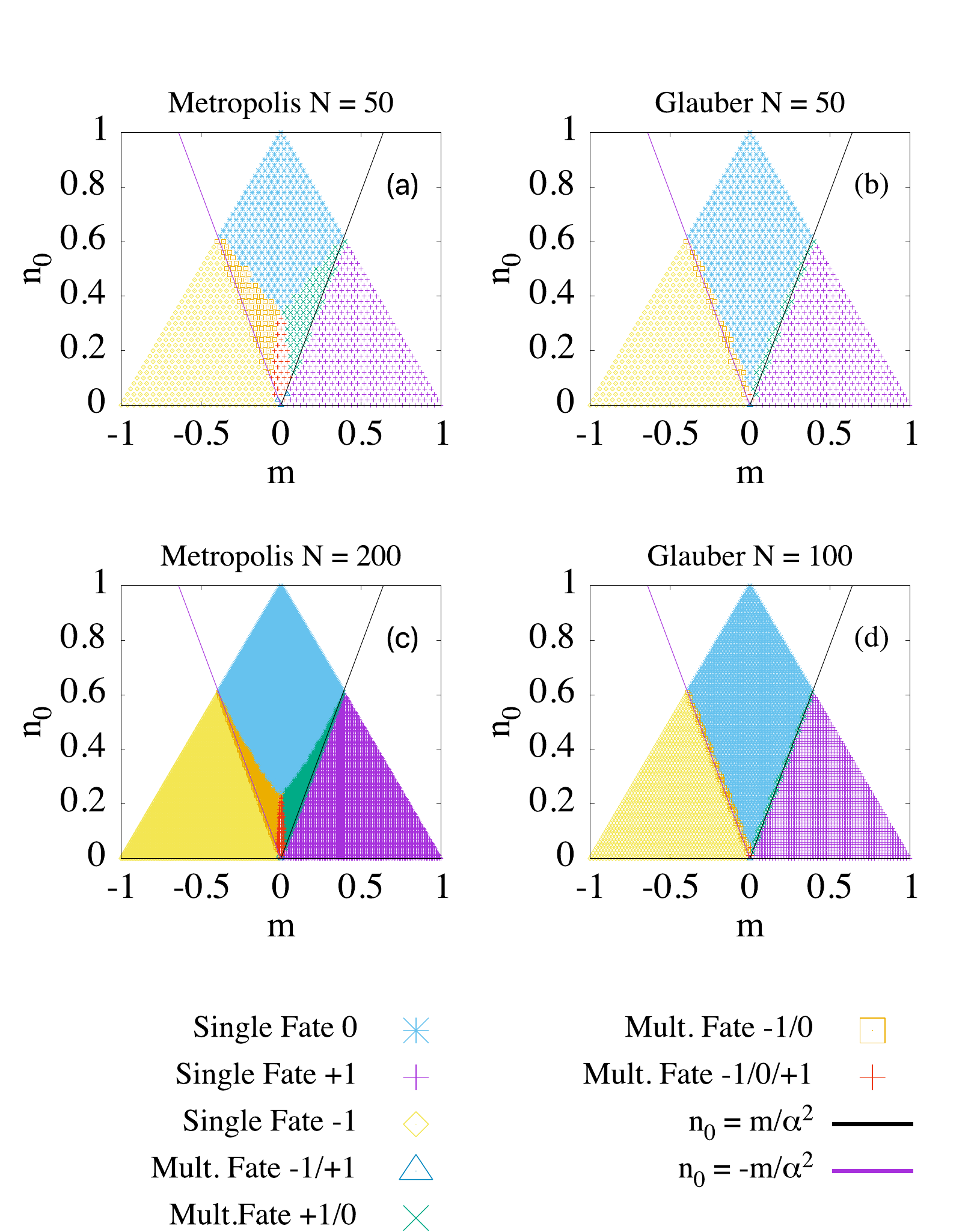}
\caption{\small{Basins of attraction at $T = 0$ and $\alpha = 0.8$ on the FC graph. (a) $N=50$, Metropolis dynamics; (b) $N=50$, Glauber dynamics; (c) $N=200$, Metropolis dynamics; (d) $N=100$, Glauber dynamics. The lines are separatrices. Symbols at each point indicate the fate of the system starting from that point. Results are obtained with $1000$ MC runs for each point 
above the separatrix, and $40$ for each point below it.}}
\label{basins}
\end{figure}

We verified the above predictions by performing repeated MC simulations with the  Metropolis dynamics at $T=0$ and
estimating, for every possible initial macrostate,  the probability
to reach each of the three aborbing states. 
In Fig.~\ref{basins} (left column) we display our results for $\alpha=0.8$. 
Indeed, points below the separatrices end up in 
the $m=\pm 1$ absorbing states, while above the separatrices we observe 
regions of ``mixed fate'' points, which we define as those
with a probability larger than 1.5\% to end up in a state different than their ``natural''
absorbing state (for example, points displayed in green can end up in $m=1$ instead of $n_0=1$).
The width of the mixed fate regions decreases with $N$,
approximately as $1/N$ as implied by the previous argument. We also verified
 that the probability of a point above the separatrix to end up in $m=1$ 
decreases exponentially with its distance from the separatrix (not shown).
An analogous analysis for the Glauber dynamics shows that, since there are fewer allowed downhill moves,
the probability to cross the separatrix is even smaller and thus the mixed-fate region is thinner,
as confirmed numerically in Fig.~\ref{basins} (right column). 

It is also interesting to observe that a deterministic downhill dynamics that follows
the direction opposite to the gradient of $H$, $-\nabla H = N (m, \alpha^2 n_0)$, also has 
the same separatrices.

From a social perspective, the above results show that for $\alpha < 1$, when we start from 
equiprobable opinions ($m=0, n_0=1/3$), the population evolves towards neutrality ($n_0=1$),
despite the optimal configurations are the polarized states. As we show below,
social agitation is necessary in order to overcome the energy barrier and reach optimality.

\subsection{Finite-temperature dynamics}

We performed MC simulations at $T>0$ with the Glauber dynamics with $N=500$ agents, again starting from a random configuration. 
When $\alpha<\alpha_{tc}$ or $\alpha>1$, the system is able to equilibrate at all temperatures $T>0.1$
in less than $10^3$ Monte Carlo steps (MCS), where 1 MCS = $N$ elementary moves. Our MC estimates for $\la m\ra$ and $\la n_0\ra$ agree with the exact mean-field results displayed in Fig.~\ref{fig:MF}, with negligible finite-size effects. 

For $\alpha_{tc}< \alpha < 1$, at temperatures below the phase boundary in
Fig.~\ref{fully_phase}, we expect metastability effects due to the 
discontinuous transition: starting from a random configuration, the system gets trapped in a region
of the configuration space with $m\simeq 0$ for a time that diverges exponentially with 
$N$ and with $1/T$.

\begin{figure}[h!]
\centering
\includegraphics[width=\columnwidth]{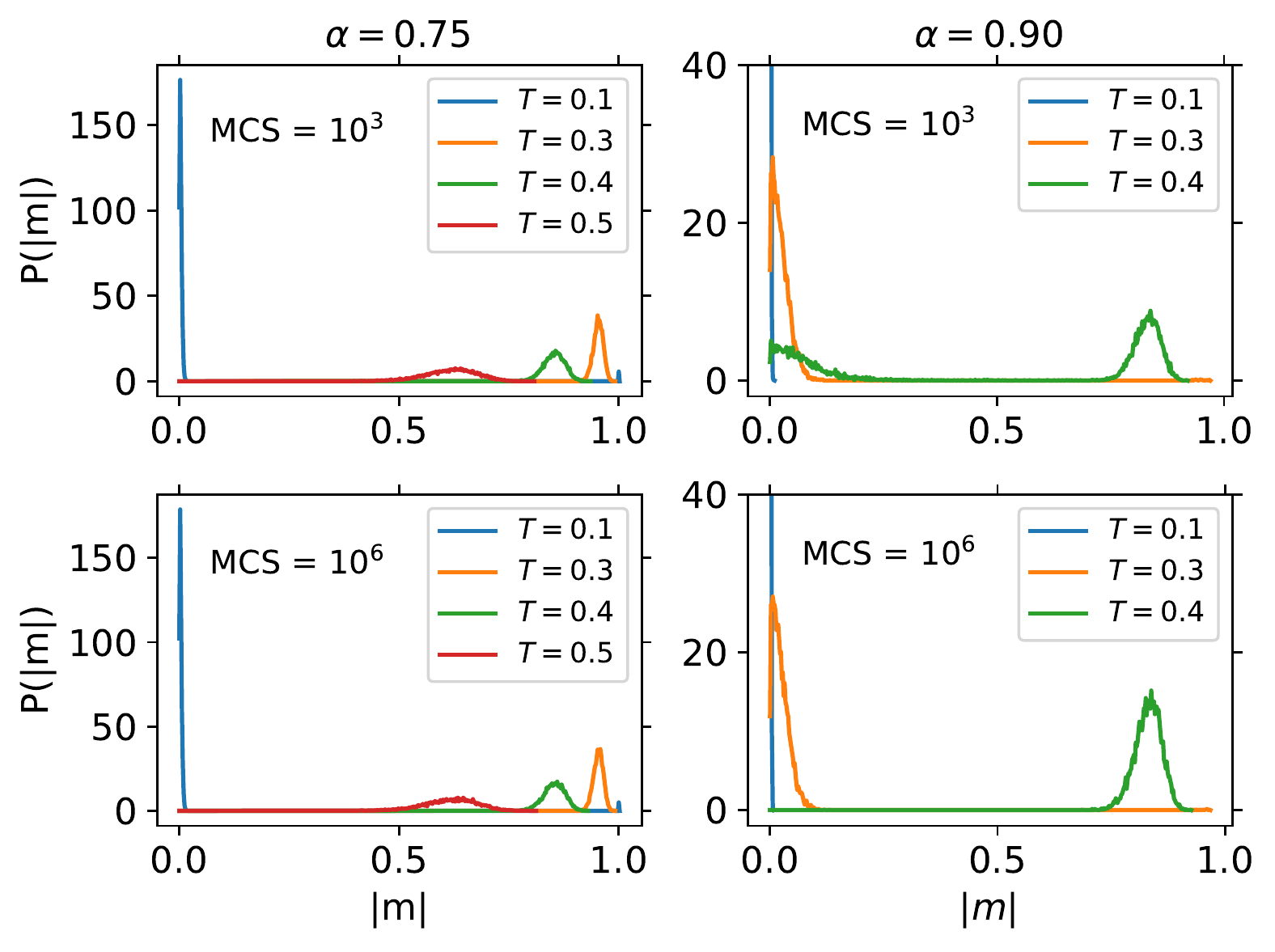}
\caption{\small{Probability distribution of the magnetization for two values of $\alpha$ below (left columb) and above (right column) the tricritical point for different temperatures obtained with Glauber MC simulations with $N=500$ agents and $10^4$ repetitions. The first row show results for a computational time of $10^3$ MCS and the second row for $10^6$ MCS.}}
\label{fig:histogram}
\end{figure}

This is confirmed by inspecting the probability distribution of the magnetization obtained from many MC runs, shown in Fig.~\ref{fig:histogram}.
The right column shows results for $\alpha=0.9$, which is well into the discontinuous region and for which the transition temperature is
$T_d = 0.47$. We see that 
for $T=0.4$, after $10^3$ MCS the distribution is bimodal, with one peak at 
the mean-field equilibrium magnetization $\langle m\ra=0.83$, and the other near $m=0$, corresponding to runs stuck in the metastable state. After $10^6$ MCS, the system is able to overcome the free-energy barrier and the peak at $m=0$ disappears. 
In contrast, for $T\leq 0.3$ even after $10^6$ MCS the distribution remains peaked around $m=0$, while the equilibrium value is close to $m=1$. This shows that almost all agents remain in the neutral state (similarly to the $T=0$ case analyzed in the 
subsection \ref{sec:T0FC}), and social agitation is not enough to overcome the free energy barrier towards polarized consensus.

This contrasts with the results for $\alpha=0.75$, shown in the left column of Fig.~\ref{fig:histogram}, 
at which the continuous transition takes place at $T_c=0.56$. We now see that for $T\geq 0.3$, already after $10^3$ MCS the distribution is peaked around the mean-field equilibrium value 
($\la m\ra = 0.95, 0.85, 0.63$ for $T=0.3,0.4,0.5$ respectively), which shows that in this case thermal fluctuations are able to bring the system towards consensus. Only at $T=0.1$ we observe that the system is more likely
to reach $m\simeq 0$ than the equilibrium value $\la m \ra \simeq 1$, even after $10^6$ MCS. 
At such low temperature, the probability to flip a polarized agent surrounded by agents of opposite sign is exponentially small.

%	

%%%%%%%%%%%%%%%%%%%%%%%%%%%%%%%%%%%%%%%%%%%%%%%%%%%%%%%%%%
\section{One-dimensional lattice} \label{One}
%%%%%%%%%%%%%%%%%%%%%%%%%%%%%%%%%%%%%%%%%%%%%%%%%%%%%%%%%

We now study the model on a 1D chain in which each agent interacts only with its
two nearest neighbors. 
The main interest of this case is that it is exactly solvable and it is the furthest from the 
FC graph. In Appendix D we solve the model 
using the transfer matrix method, adopting periodic boundary conditions (it is straightforward
to consider other boundary conditions, such as open or free, and the results for large $N$ are not affected
by the choice). In the large-$N$ limit, we obtain the free energy per site
\begin{eqnarray}
\nonumber
f(\beta,\alpha) &=& -\frac{1}{\beta}\ln \left[ \cosh \beta + e^{\beta \alpha^{2}}/2  \right.
\\  &+&  
\left. \sqrt{(\cosh \beta - e^{\beta \alpha^2}/2)^2 + 2} \,\,\right] .
\label{eq:1Df}
\end{eqnarray}
This is an analytic function of its arguments, thus it does not display any 
finite-temperature phase transition, as expected in any 1D models with short-range interactions.
Only in the zero-temperature limit there is a phase transition at $\alpha = 1$.
In fact, the limit of  $f(\beta,\alpha)$ for $\beta\to\infty$ gives the ground-state energy per site,
which is $-1$  for $\alpha\leq 1$, and $-\alpha^2$ for $\alpha > 1$.

The transfer-matrix solution also provides an explicit expression for the correlation function $C(r)=\langle \sigma_i \sigma_{i+r}\rangle$. As shown in Appendix D, in the $N\to\infty$ limit we obtain $C(r) = c_0 \exp[-r / \xi(\beta)]$ where $c_0$ is a 
prefactor that depends on $\beta$ and $\alpha$. The correlation length $\xi(\beta)$ is
given by
\beqn
\frac{1}{\xi(\beta)} &=& \ln\left(\cosh \beta + e^{\beta \alpha^2}/2\right. \label{xi}
\\ \nonumber & +& \left. \sqrt{( \cosh \beta - e^{\beta \alpha^2}/2)^2 + 2}\right) - \ln 2 \sinh \beta  \,.
\label{none}
\eeqn

For $\alpha<1$ the correlation length diverges exponentially at low temperatures as
\beq
\xi(\beta) \underset{\beta \gg 1} \sim \frac{1}{4}e^{2 \beta} \,,
\label{xi_asymp}
\eeq
where the exponent stems from the fact that the energy of a domain wall between
a positively polarized domain (i.e. a segment of contiguous agents with $\sigma_i=1$)
and a negatively polarized domain is $2$.
%\footnote{Apart from the prefactor, Eq.(\ref{xi_asymp}) can also be deduced from a Peierls argument.}. 

We now use this result to deduce the behavior of the magnetization.
If $N \gg \xi$, we can consider the system as composed of $N/\xi$ independent domains of
alternating sign, each with a length of order $\xi$. The 
magnetization is thus $m \simeq \sum_{k=1}^{N/\xi} M_k / N$, where 
the domain magnetizations $M_k \simeq \pm \xi$ can be treated as independent random variables with variance of 
order $\xi^2$. From this we obtain the scaling law
\beq
\la |m|\ra^2 N \sim \xi(\beta).
\label{scaling_m_1D}
\eeq

On the other hand, when $N < \xi$, if we let the system evolve with the Glauber or Metropolis dynamics
starting from a random configuration, at infinite
time it will reach a magnetized state with $\la |m|\ra \simeq 1$. The evolution towards this state takes
place via a domain-growth process (known as coarsening in statistical physics), in
which the walls between domains of different sign perform a random walk. When two domain walls collide,
they annihilate each other, fusing the domains, so $\la |m|\ra$ will grow in time
until approaching $\simeq 1$.

\begin{figure}[h!]
\centering
\includegraphics[width=\columnwidth]{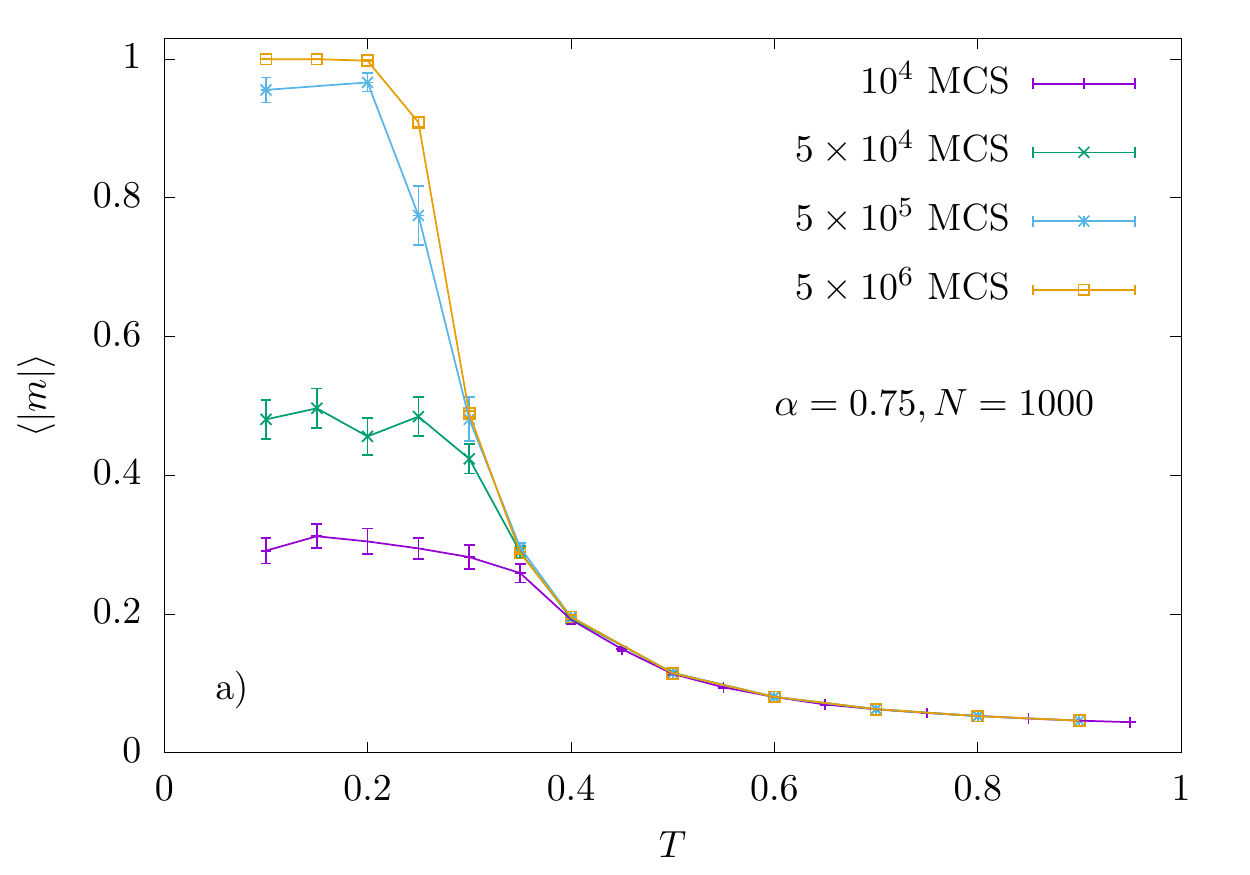}

\includegraphics[width=\columnwidth]{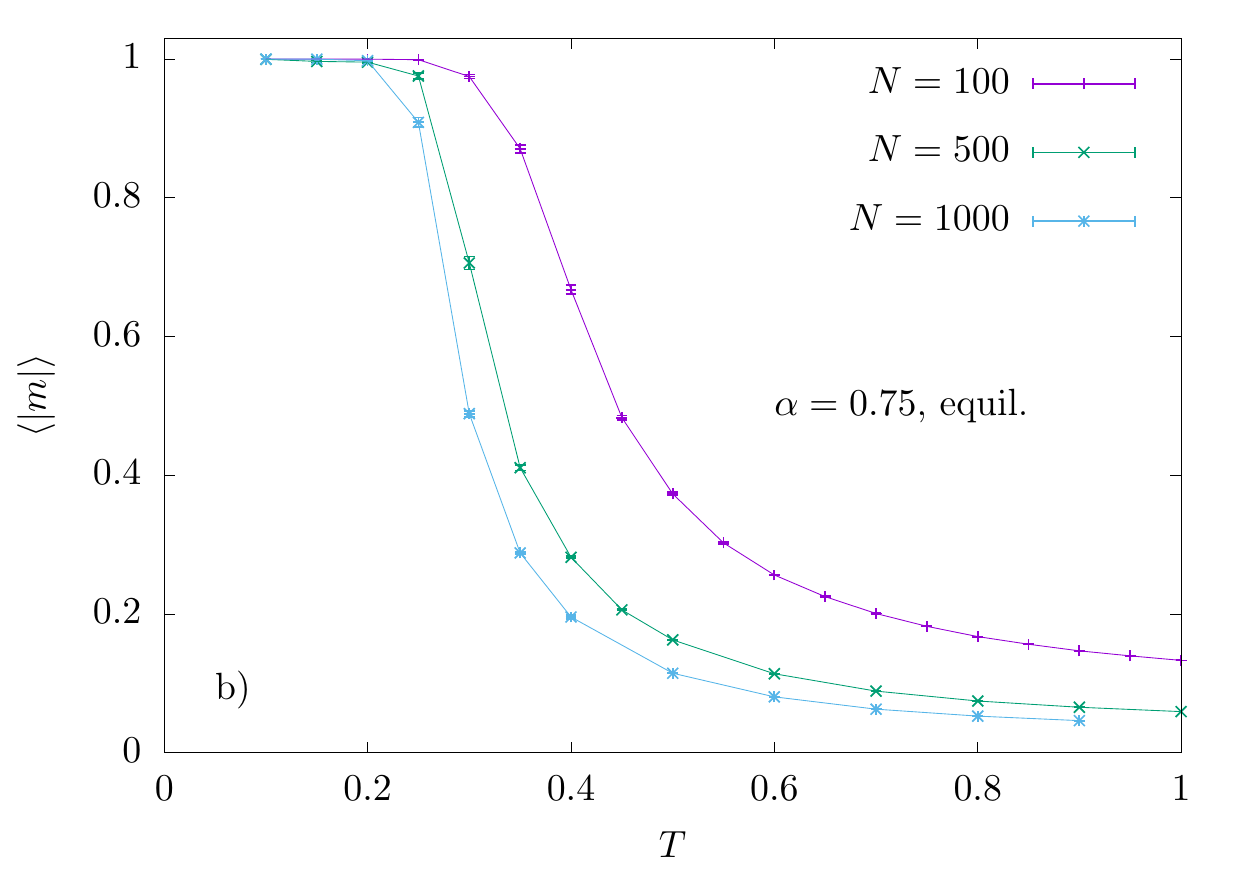}
\caption{\small{Monte Carlo results for the average absolute magnetization of the 1D chain
as a function of temperature, for $\alpha=0.75$.
For each data set, the initial 1/5 of the time series was discarded, and data were averaged
over 100-400 repetitions. Periodic boundary conditions were applied.
(a) Data for $N=1000$ and different simulations times, starting from a random configuration.
(b) Data for different $N$, for sufficiently large simulation time so that $\la|m\ra$ is equilibrated
($5\times 10^4,  5\times 10^5,  5\times 10^6$ MCS for $N=100, 500, 1000$ respectively). Lines are only a guide to the eye.
}}
\label{magn1D}
\end{figure}

The results of our MC simulations agree very well with the above
predictions. Fig.~\ref{magn1D}(a) shows, for $N=1000$ and $\alpha=0.75$, $\la |m|\ra$ as a function 
of temperature, for different simulation times. Indeed one can observe that at high $T$
the data are in thermal equilibrium, while at low $T$, $\la |m|\ra$ grows with time until 
approaching unity.

\begin{figure}[h!]
\centering
\includegraphics[width=\columnwidth]{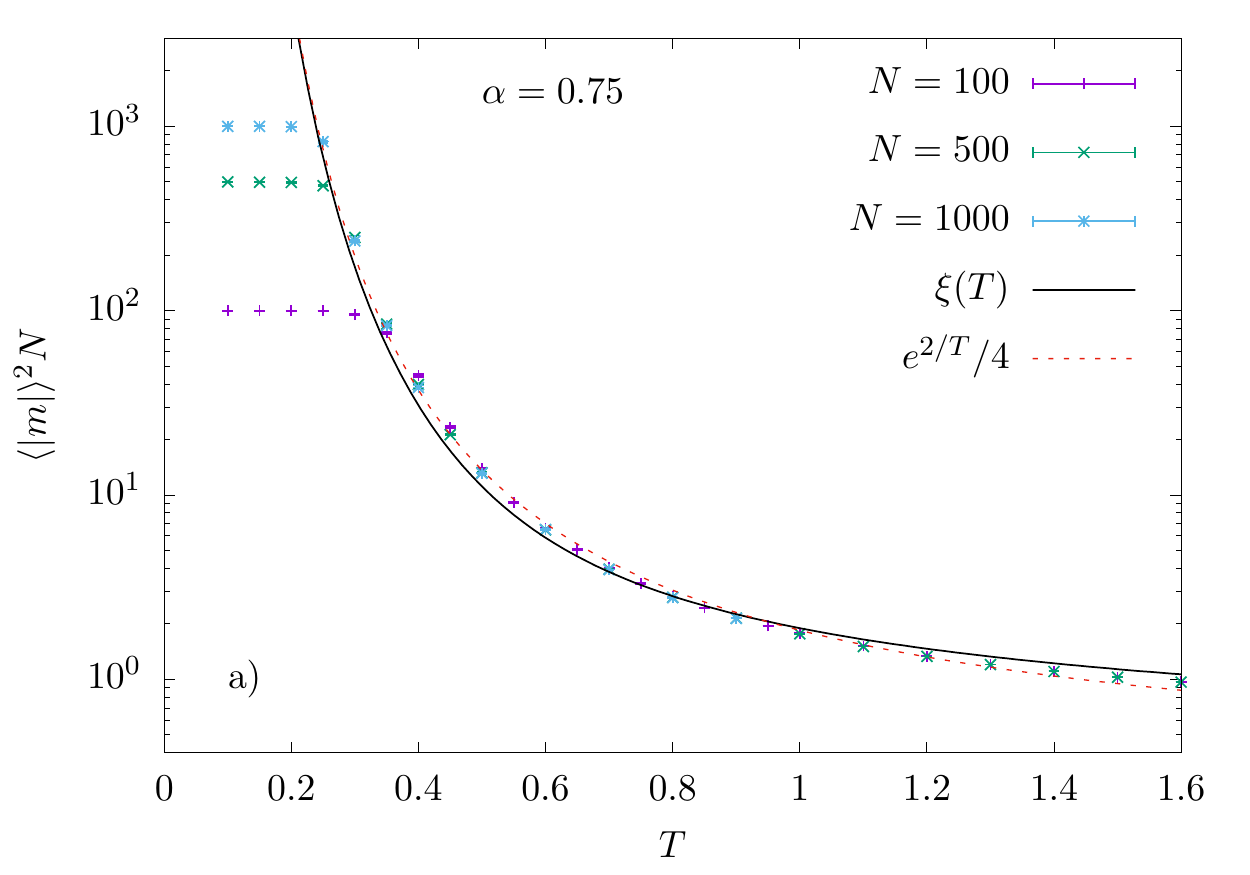}
\includegraphics[width=\columnwidth]{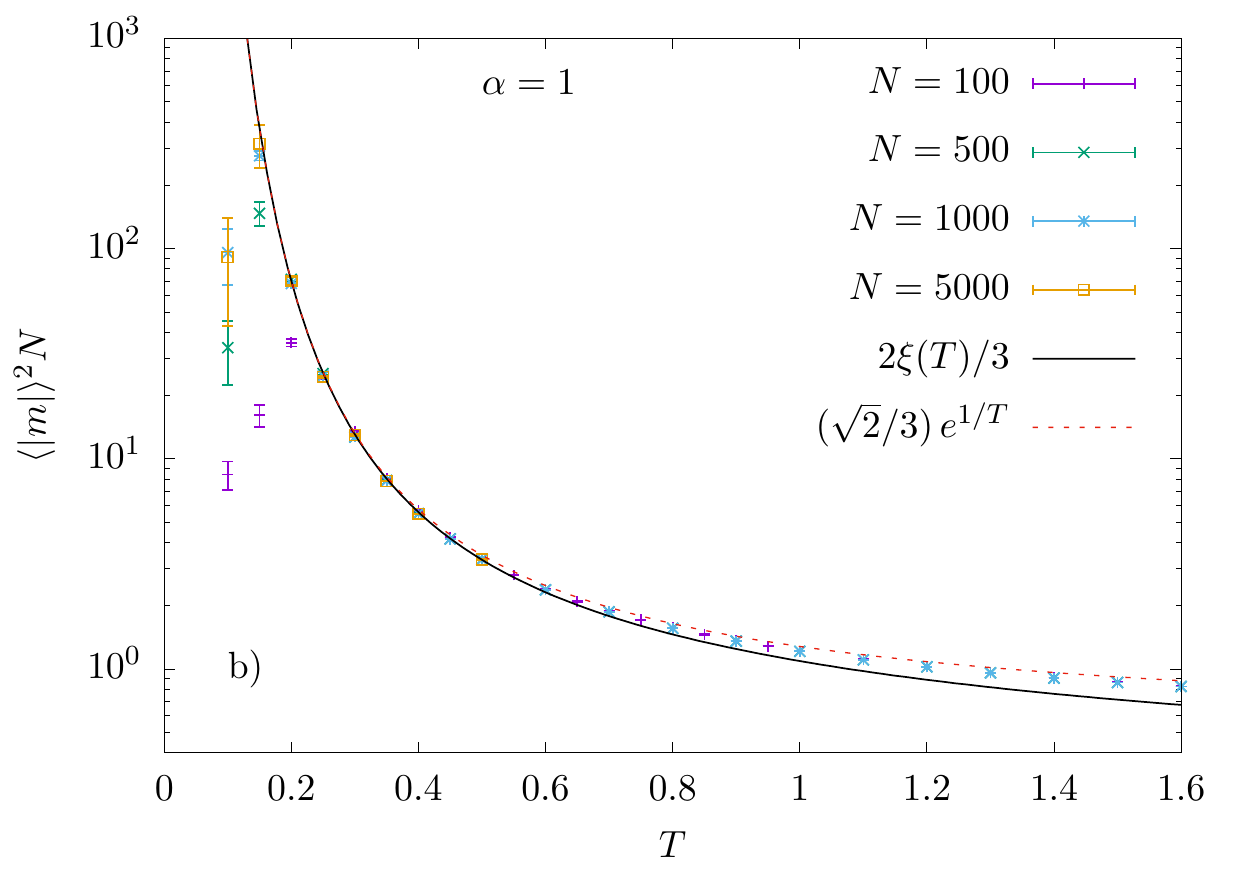}
\caption{\small{(a) Scaling of the average absolute magnetization of the 1D chain
   for $\alpha < 1$. Data are the same as
 in Fig.~\ref{magn1D}(b),
 rescaled according to Eq.(\ref{scaling_m_1D}). The curves
show the full expression of the correlation length Eq.(\ref{xi}) and the
asymptotic behavior Eq.(\ref{xi_asymp}).
(b) Same as a) but for $\alpha=1$ and scaling the data according to Eq.\eqref{scaling_m_1D_a1}.
The data for $L=5000$ were obtained with $10^6$ MCS, the others with the same number
of MCS as in (a). 
The curves show the full expression of the correlation length Eq.(\ref{xi}) and the
asymptotic behavior for $\alpha=1$, Eq.(\ref{xi_asymp_a1}).}}
\label{magn1D_scaling}
\end{figure}

Fig.~\ref{magn1D}(b) shows the equilibrium 
value of the magnetization achieved after a sufficiently long time, for different $N$:
as $N$ grows, the magnetization saturates when $T$ is below the temperatures at which $\xi$ overcomes $N$.
Fig.~\ref{magn1D_scaling}(a) shows that the data for different $N$ in the regime $N\gg\xi$ can be rescaled
very well according to Eq.(\ref{scaling_m_1D}).

At the transition point $\alpha=1$, the behavior is somewhat different. In fact, the 
correlation length now diverges at low temperatures as
\beq
\xi(\beta) \underset{\beta \gg 1} \sim \frac{1}{\sqrt{2}}e^{\beta} \,,
\label{xi_asymp_a1}
\eeq
since the energy of a domain wall between neutral domains and polarized domains is $1$.
Following the same argument given above for $\alpha< 1$, the domain magnetizations can be either 
$M_k \simeq 0$ or $M_k\simeq \pm \xi$ with equal probability, and thus have variance $\simeq 2\xi^2/3$,
which gives
\beq
\la |m|\ra^2 N \sim \frac{2}{3} \xi(\beta).
\label{scaling_m_1D_a1}
\eeq
In this case, for $N < \xi$, the equilibrium magnetization saturates to a value less than $1$ as
the system can evolve towards either one of the three consensus states.

Finally, the case $\alpha > 1$ is simpler as now there is a unique ground state, thus the system evolves quickly towards the equilibrium configuration.

%%%%%%%%%%%%%%%%%%%%%%%%%%%%%%%%%%%%%%%%%%%%%%%%%%%%%%%%%%
\section{Erd\"os-R\'enyi graphs} \label{Erdos}
%%%%%%%%%%%%%%%%%%%%%%%%%%%%%%%%%%%%%%%%%%%%%%%%%%%%%%%%%%

Although the ER random graph ensemble is not representative of real-world structures, it is often used as a null model to study complex networks \cite{Newman2002}. 
We obtain the equilibrium behavior of the model (averaged over the ER ensemble)
using the annealed approximation \cite{Bianconi_2002, Dorogovtsev_2008}, 
which  consists
in applying the mean-field SCEs in Eqs.\eqref{eq:SCE_m} and \eqref{eq:SCE_n}) taking
into account that the degree $k$ has a Poisson distribution 
$P(k)=e^{-z} z^k/k!$ with average degree $z$ \cite{Erdos:1959:pmd}. 
In this way we obtain
\begin{eqnarray}
m &=\sum_{k} P(k) \frac{2 e^{\beta \alpha^2 k (n_w-1)} \sinh(\beta k m_w)}{1+2 e^{\beta \alpha^2 k (n_w-1)} \cosh(\beta k m_w)},\label{eq:ER1} \\
n &=\sum_{k} P(k) \frac{2 e^{\beta \alpha^2 k (n_w-1)} \cosh(\beta k m_w)}{1+2 e^{\beta \alpha^2 k (n_w-1)} \cosh(\beta k m_w)} .
\label{eq:ER2}
\end{eqnarray}
where $m_w  =\sum_j k_j m_j/zN$ and  $n_w = \sum_j k_j n_j/zN$  are weighted order parameters that satisfy two coupled SCEs (see Appendix C for details).

\subsection{Phase transition}
\begin{figure}[h!]
\centering
\includegraphics[width=\columnwidth]{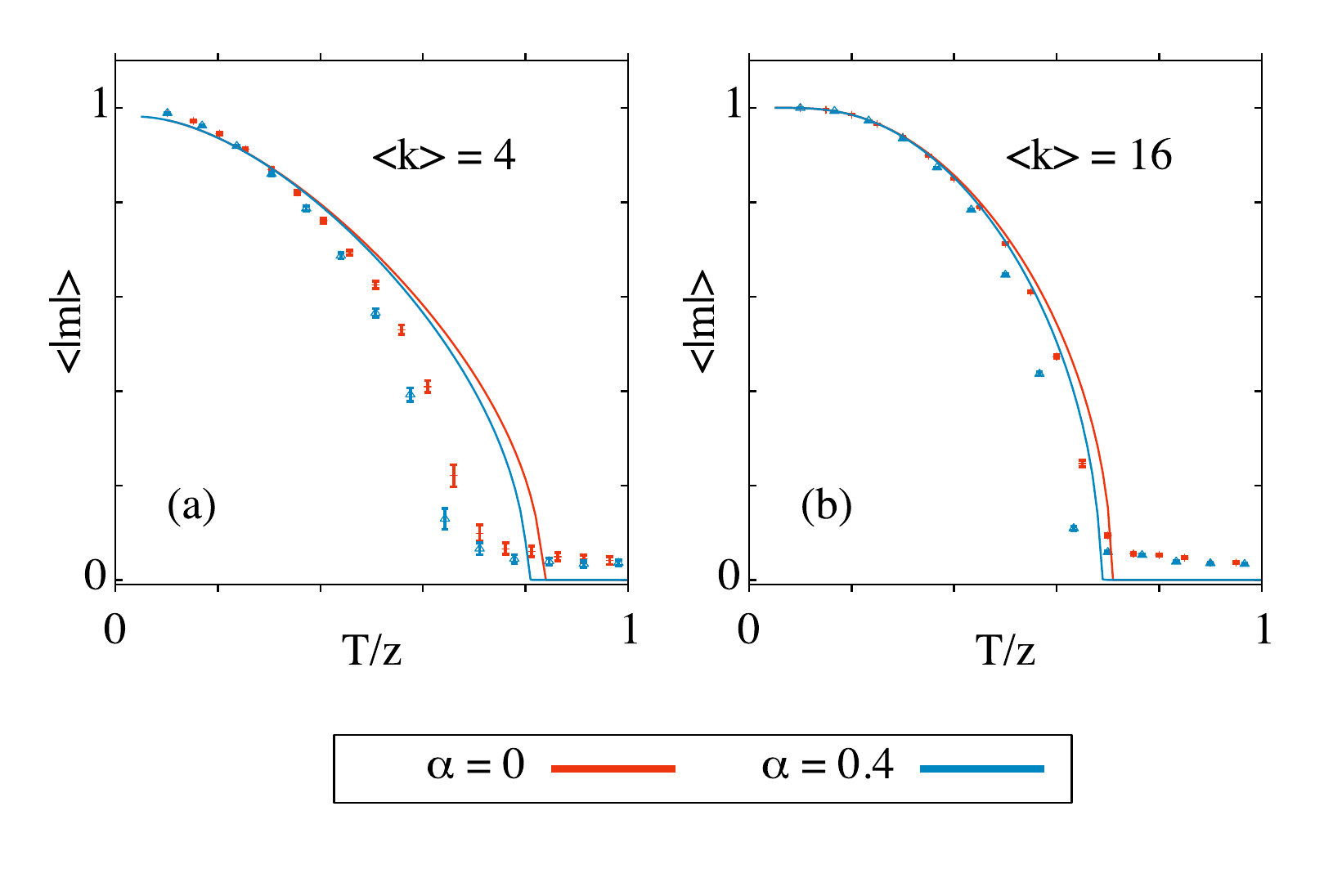}
\caption{\small{Magnetization vs. temperature on ER graphs of average degree $z = 4$ (a) and $z = 16$ (b) for $\alpha = 0$ and $\alpha = 0.4$}. The lines corresponds to the numerical solution in the annealed approximation, while the bars are the results of MC simulations with $10^4$ MCS, averaged on 100 realizations of different ER graphs with $N=1000$ agents.}
\label{ER}
\end{figure}
Also in this case one expects in general a tricritical point separating a line of continuous transitions for
low $\alpha$ from a line of discontinuous transitions at large $\alpha$, as in
the Blume-Capel model on ER graphs \cite{Leone}.
We determined the location of the continuous transition by solving numerically the SCEs for $m_w, n_w$ using a Broyden first Jacobian approximation from the 
optimize root of SciPy.
Fig.~\ref{ER} shows the comparison between the numerical solution and MC simulations for $\alpha = 0$ and $0.4$.
For low connectivity (panel (a)), we found a significant discrepancy between the annealed approximation and the simulations, while for high connectivity (panel (b)) we obtain a better agreement. 
\begin{figure}[h!]
\centering
\includegraphics[width=\columnwidth]{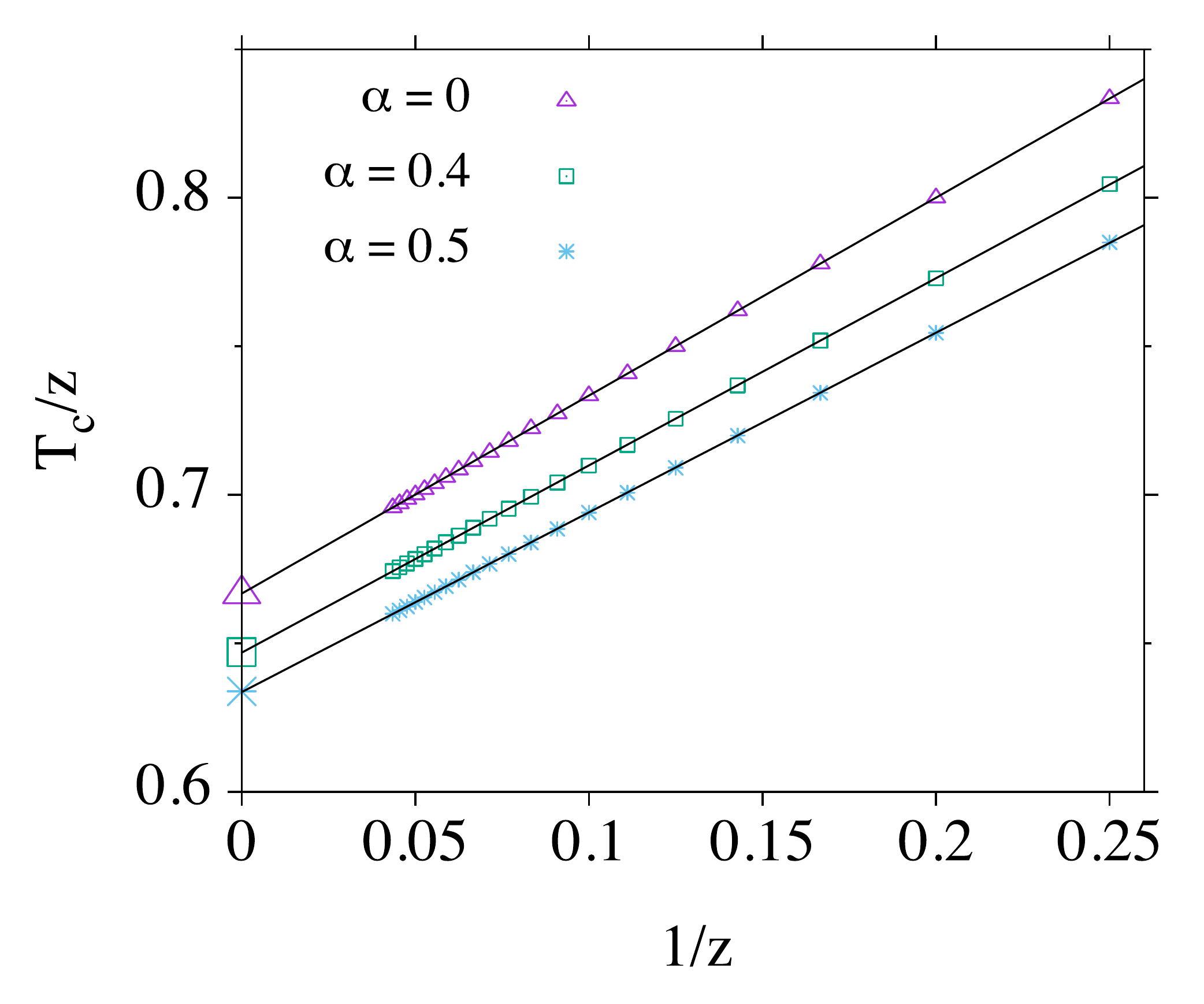}
\caption{\small{Critical temperature vs. $1/z$ for different values of $\alpha$, calculated by solving numerically the system of equations (\ref{eq:ER1}), (\ref{eq:ER2}). Mean-field results are represented by bigger points at $z\to \infty$.}}
\label{ER_Tc}
\end{figure}

Fig.~\ref{ER_Tc} shows that the critical temperature, normalized by $z$,  converges asymptotically for large $z$ to that of the FC graph. 
The critical temperature decreases with $\alpha$ for any $\langle k \rangle$, as in the mean-field solution. 

The putative tricritical point and the
discontinuous transition can in principle be located from a Taylor expansion
in $m$ of the annealed SCEs. We have have not attempted this, but we note that it
is more complicated than in the Blume-Capel model \cite{Leone}, in which the 
SCE $m$ decouples from $n$.

\subsection{Zero-temperature dynamics}

We performed MC simulations with the Metropolis and Glauber dynamics at $T=0$,
starting from a random configuration as in the FC graph. 
We consider graphs with only one connected component, created by 
generating ER graphs with the Python $networkx$ library and then randomly adding and/or subtracting agents and 
links until the desired $N$ and $z=\langle k \rangle$ are reached, preserving the degree distribution of the original graph. 

\begin{figure}[h!]
\centering
\includegraphics[width=\columnwidth]{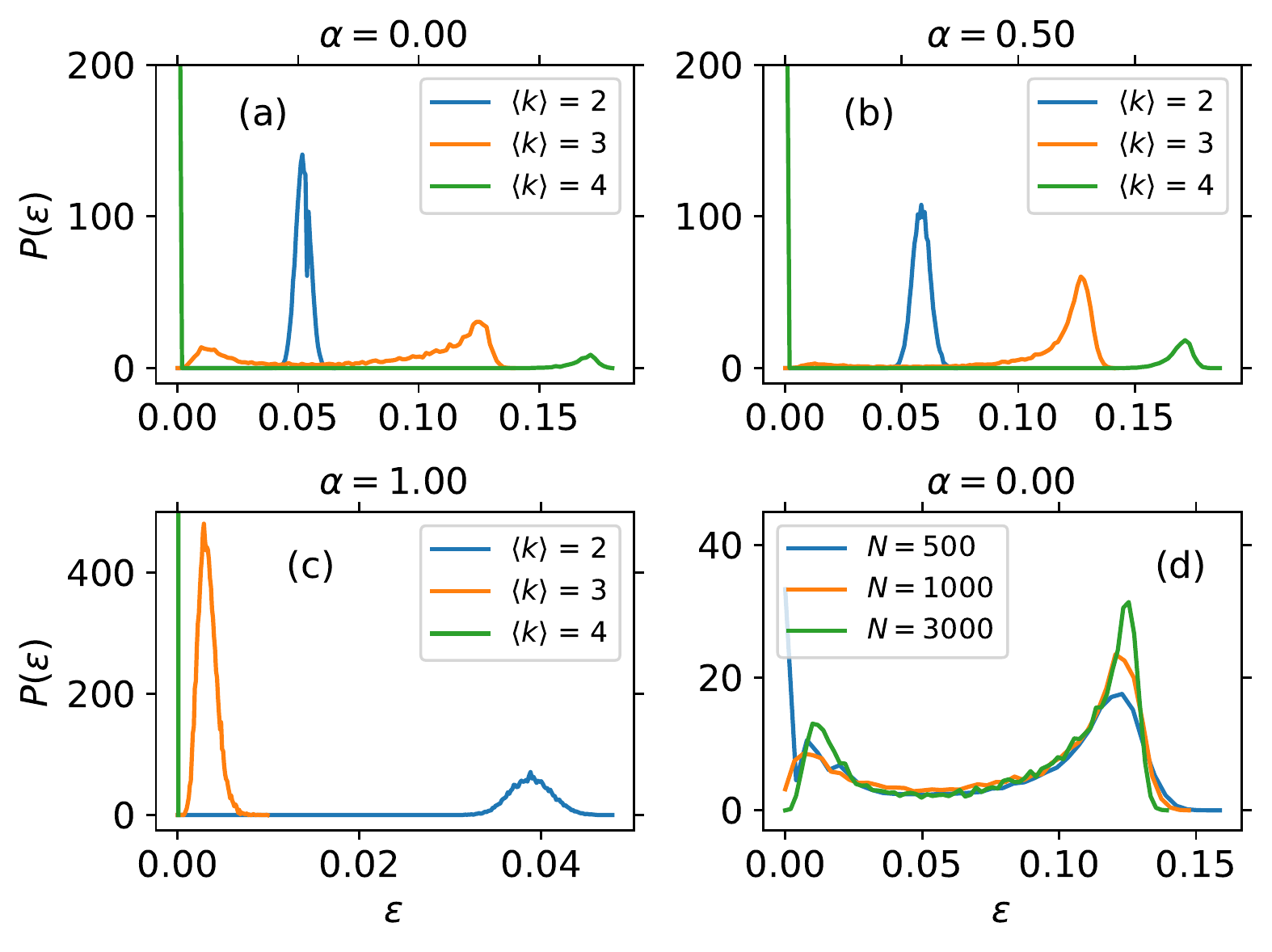}
\caption{\small{Probability distribution function of the residual energy $\epsilon$, for the steady state reached through Glauber dynamics on a ER graph with $N = 3000$ and $z=\langle k \rangle = 2, 3, 4$ for (a) $\alpha = 0$, (b) $\alpha = 0.5$, and (c) $\alpha = 1$. (d) Comparison of the probability distribution function for $\langle k \rangle = 3$ and different values of $N$. Data are obtained from $10^5$ different random initial conditions on a single realization of the network for $N=500$ and $2\cdot 10^4$ different random initial conditions for $N=1000, 3000$.}} 
\label{ER_res}
\end{figure}

We perform many runs for each graph, letting the system relax until it reaches a steady state in which the energy no longer changes.
In this way we collect the probability distribution $P(\epsilon)$ of the residual energy, defined as the normalized difference between 
the energy of the steady state reached in a given run and the ground-state energy of the graph, $\epsilon = (E_{steady} - E_{0})/N z$. 
Fig.~\ref{ER_res} shows the results obtained with the Glauber dynamics for $z\leq 4$ and different values of $\alpha$.
For $z=2$, in all cases $P(\epsilon)$ has a single peak at
$\epsilon >0$, indicating that the system never reaches the ground state as it gets stuck in a manifold of excited isoenergetic configurations,
the energy of which changes from run to run. The same behavior persists for $\alpha > 1$ (see Appendix E for $N=500$).

\begin{figure}[h!]
\centering
\includegraphics[width=\columnwidth]{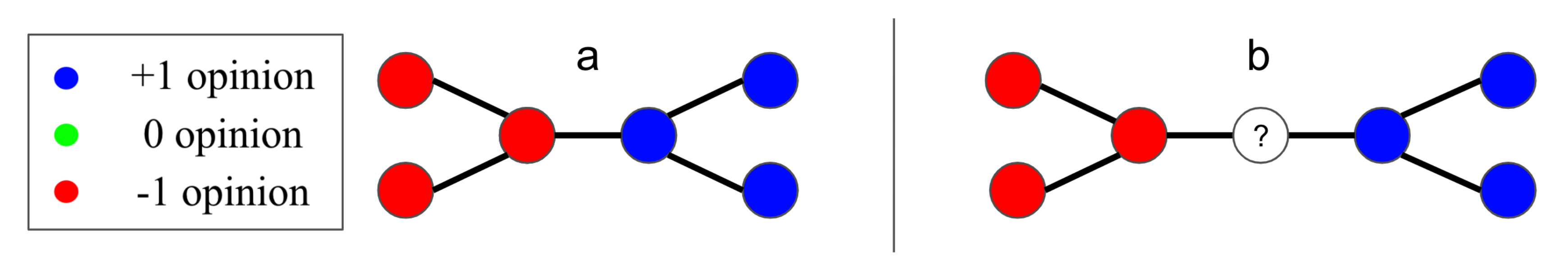}
\caption{\small{Example of subgraphs with a stuck spin configuration, despite not being compatible with the ground state.}}
\label{blinkers}
\end{figure}

A similar behavior has been observed before for the Ising model \cite{Baek2012},
and H\"aggstr\"om \cite{Hggstrm2002} showed rigorously that it
stems from the existence of an extensive number of 
subgraphs in which some nodes are frozen 
(i.e. they cannot change state without increasing the energy) in an excited 
configuration. In addition, some nodes are ``blinkers'', namely
they can change state forever without changing the energy, thus the system
gets trapped in a manifold of configurations at constant energy above the ground state. As an illustration, the two central nodes of Fig.~\ref{blinkers} a) are frozen, while 
the middle node in Fig.~\ref{blinkers} b) is a blinker. 

Upon increasing $z$, the number of frozen nodes and blinkers decreases, 
hence so does the probability to get stuck in dynamical traps.
For $z=3$, as we increase $\alpha$ we see that the distribution is bimodal for $\alpha\leq 0.75$,
with both peaks at $\epsilon > 0$ (the ground state is never reached), then 
the peak closer to $\epsilon=0$ disappears for $0.85 \leq \alpha \leq 0.95$, 
and finally the distribution becomes unimodal with a finite weigth at $\epsilon=0$ for $\alpha > 1$ (see Appendix E).
Note that these dynamical traps are not a finite-size effect: as shown 
in Fig.~\ref{ER_res} d) for  $z=3$ and $\alpha = 0$, the peak near $\epsilon=0$ decreases with the system size,
and the one at larger $\epsilon$ grows.

For $z=4$, we observe a large probability $P(0)$ to reach the ground state for all values of $\alpha$.
For $\alpha<1$ there is still a peak at $\epsilon>0$
but it is significantly smaller than that for $z=3$. 
Notice that for both $z=3$ and $z=4$ the peak at large $\epsilon$ increases significantly in the range 
$0.85\leq \alpha\leq 0.95$:
it is an interesing question whether this is related to existence of a tricritical point.

Finally, for the Metropolis dynamics we obtain very similar results to the ones just discussed.
A comparison between the two 
dynamics for several values of $\alpha$,  $z$ and $N$ is shown in the Appendix E.

The above results show that the relaxational $T=0$ dynamics on ER graphs 
is quite different from the FC case in which, as shown 
in Section \ref{FCG}, for large $N$ the system prepared in a random configuration always reaches 
the ground state if $\alpha>1$  (namely the absorbing state $n_0=1$), and never reaches it 
if $\alpha < 1$. In ER graphs, instead, the system is able to reach
the ground state for all $\alpha$, except for small values of $z$.


\section{Conclusions} \label{conclusions}
We have studied a three-state Hamiltonian model aimed at understanding consensus formation in the presence of neutral agents between opposite extremes.
The tendency of people to align their opinions with those of their friends is captured by a pairwise Ising-like
interaction including a neutral orientation weighted
 by a neutrality parameter $\alpha$. This tendency is opposed 
by individual thinking, social agitation and other factors, here collectively represented as a temperature.

Regardless of the network type, at high temperatures the agents do not feel the influence of their peers and evolve quickly towards a disordered state, that can be interpreted as social unrest.
At low temperature, for $\alpha>1$ the system reaches a configuration dominated by neutral agents
 on all networks, while for $\alpha<1$ we observe different 
behaviors depending on the network type, and on the value of $\alpha$, as
we summarize below.

On the FC graph, the equilibrium phase diagram of the model exhibits a phase boundary
for $\alpha < 1$ between a low-temperature phase, in which one of the polarized opinions prevail, and the high temperature disordered phase. At temperatures just above the transition, we observe a majority of neutral agents and equal-sized minorities of extremists of both signs. Upon crossing the phase boundary by decreasing the temperature, 
the fraction of polarized agents increases continuously from zero when $\alpha$ is below the tricritical value $\alpha_{tc}$, 
while it jumps discontinuously to a non-zero value
when $\alpha_{tc}<\alpha<1$. We also found that in the absence of social agitation, starting from a random configuration, the population is exponentially more likely to get stuck in neutral consensus than to reach
the optimal polarized consensus. Moderate social agitation allows the system  to reach polarized consensus for $\alpha < \alpha_{tc}$, while for $\alpha<\alpha_{tc}<1$ the barriers to achieve equilibrium are much higher due to metastability.

It is interesting to compare our results with those of other three-state models on the FC graph.
Vazquez and Redner \cite{Vazquez2004} studied
a kinetic model in which pairs of agents interact stochastically.
Their model exhibits the same three absorbing states, but the largest basin of attraction is 
that of an absorbing boundary on the bottom line of the triangle (see Fig.\ref{flow}), 
representing frozen mixtures of oppositely polarized agents.  
A large bipolarized region was also found by Balenzuela et al.\cite{Balenzuela_2015},   who proposed a kinetic model in which agents hold a continuous spectrum of convictions, which is partitioned in three states according to some thresholds. The pair interaction is such that oppositely polarized agents tends to increase their polarization, leading to a phase transition from a neutral to either a bipolarized or a polarized population.
Svenkeson and Swami \cite{Svenkeson2015} considered a pair dynamics, in which polarized agents can become neutral and viceversa, with a temperature coupled to the instantaneous magnetization. They find a transition of the Ising type without a tricritical point. 

Our results for the one-dimensional chain, which obviously does not display a phase transition,
show that the evolution towards consensus takes place via the growth of domains of contiguous like-minded polarized agents, until one domain takes over the whole population.

Finally, in ER random graphs, which are somewhat closer to a real population in which agents have a finite number of contacts, we observe a finite-temperature transition 
to a polarized phase, analogous to that of the FC graph. However, unlike in the FC graph, for large enough connectivities the system is always able to reach the polarized state from a random start, while for low connectivities it gets stuck in dynamical traps due to frozen nodes. 

The behavior of the model on networks that more realistically represent human connection patterns, either synthetic or extracted from real data, will be considered elsewhere.

%%%%%%%%%%%%%%%%%%%%%%%%%%%%%%%%%%%%%%%%%%%%%%%%%%%%%%%%%%%%%%%%%%%%%%%%

\vspace*{0.5cm}
\begin{acknowledgments}
We acknowledge financial support from MINECO via Project No. PGC2018-094754-B-C22 (MINECO/FEDER,UE) and Generalitat de Catalunya via Grants No. 2017SGR341 and 2017SGR1614. We are pleased to thank Conrad P\'erez Vicente for discussions.

\end{acknowledgments}

%%%%%%%%%%%%%%%%%%%%%%%%%%%%%%%%%%%%%%%%%%%%%%%%%%%%%%%%%%%%%%%%%%%%%%%%%
%%%%%%%%%%%%%%%%%%%%%%%%%%%%%%%%%%%%%%%%%%%%%%%%%%%%%%%%%%%%%%%%%%%%%%%%

%\bibliographystyle{apsrev4-2}
%\bibliographystyle{ieeetr}
%\bibliographystyle{plain}

%\bibliographystyle{ieeetran}
%\bibliographystyle{physrevstyle}
%\bibliography{mybib_abbrv}
%

%%%%%%%%%%%%%%%%%%%%%%%%%%%%%%%%%%%%%%%%%%%%%%%%%%%%%%%%%%%%%%%%%%%%%%%%%
%%%%%%%%%%%%%%%%%%%%%%%%%%%%%%%%%%%%%%%%%%%%%%%%%%%%%%%%%%%%%%%%%%%%%%%%
\onecolumngrid

\begin{appendix}

%%%%%%%%%%%%%%%%%%%%%%%%%%%%%%%%%%%%%%%%%%%%%%%%%%%%%%%%%%%%%%%%%%%%%%%%%
%%%%%%%%%%%%%%%%%%%%%%%%%%%%%%%%%%%%%%%%%%%%%%%%%%%%%%%%%%%%%%%%%%%%%%%%
\section{Stochastic dynamics}
\renewcommand{\theequation}{A-\arabic{equation}}

The following sequences decribe one elementary move of the stochastic dynamics 
used for our simulations.

\bigskip 
\underline{\textit{Metropolis dynamics}}
\bigskip 

\textbf{Step 1}: Pick one agent $i$ at random with uniform probability among the $N$ agents.

\textbf{Step 2}: Propose a random opinion change with probability 1/2 
for the agent $i$ and calculate the energy difference $\Delta H$ between the
new and old configuration.

\textbf{Step 3}: Accept the change with probability $\min \{1, e^{-\beta \Delta H} \}$,
otherwise remain in the current state.

\bigskip 
\underline{\textit{Glauber dynamics}}
\bigskip 

\textbf{Step 1}: Pick one agent $i$ at random with uniform probability among the $N$ agents.

\textbf{Step 2}: Calculate the energy differences $\Delta H_{1}$ and $\Delta H_{2}$ between
the new and old configuration, for the two possible opinion changes for the agent $i$.

\textbf{Step 3}: Accept either of the two possible changes with probability 
$P_{k} = w_k/(1+w_1+w_2)$, where $w_k = e^{-\beta \Delta H_{k}}$ for  $k=1,2$,  
or remain in the current state with  probability $1-P_1-P_2$.

%%%%%%%%%%%%%%%%%%%%%%%%%%%%%%%%%%%%%%%%%%%%%%%%%%%%%%%%%%%%%%%%%%%%%%%%%
\section{Mean-field solution}
\renewcommand{\theequation}{B-\arabic{equation}}
\label{app:ch1}

We derive here the mean-field solution of the model, which is exact for the FC case in the large $N$ limit. Consider first a generic graph in which node $i$ has degree $k_i$. The mean-field approximation consists in assuming that 
the spins $\sigma_i$ are independent random variables
that can take the values $\sigma_i=\pm 1$ with probabilities $p_i^\pm$ 
and $\sigma_i=0$ with probability $1 -p_i^+ - p_i^-$. 
The Hamiltonian in Eq.\eqref{eq:hamiltonian_2} takes then the mean-field form
\beq
H_{MF} = - \sum_{\langle i,j\rangle} [m_i m_j + \alpha^2 (1-n_i)(1-n_j)]
\eeq
where $m_i = p_i^+ - p_i^-$ are the local magnetizations and $n_i = p_i^+ + p_i^-$
is the probability that $\sigma_i$ is different than zero.
The mean-field entropy is given  by
\beqn
S_{MF} &=& - \sum_i \left[ p_i^+ \ln p_i^+ + p_i^- \ln p_i^- + (1-n_i) \ln (1-n_i) \right]
\nonumber \\&=& -\sum_i \left[ \frac{n_i+m_i}{2} \ln \frac{n_i+m_i}{2} + \frac{n_i-m_i}{2} \ln \frac{n_i-m_i}{2} + (1-n_i) \ln (1-n_i) \right].
\eeqn
The equilibrium values of $m=\{m_i\}, n=\{n_i\}$ are those that minimize the free-energy function 
${\cal L}(m,n,\beta) \equiv (H_{MF}- \beta^{-1} S_{MF})/N$ and thus satisfy the system
of $2N$ coupled equations
\beqn
\frac{\partial {\cal L}}{\partial m_i} &=& - \sum_{j\in V(i)} m_j + \frac{1}{\beta} \mbox{arctanh} \frac{m_i}{n_i} =0 \\
\frac{\partial {\cal L}}{\partial n_i} &=& \alpha^2 \sum_{j\in V(i)} (1-n_j) + \frac{1}{2 \beta} 
\ln \frac{n_i^2 - m_i^2}{4 (1-n_i)^2} = 0
\eeqn
where $V(i)$ denotes the set of nodes connected to node $i$. The above equations can be written as
\begin{subequations}
\begin{align}
m_i &= \frac{2 e^{\beta \alpha^2 \sum_{j\in V(i)} (n_j-1)} \sinh(\beta \sum_{j\in V(i)} m_j)}{1+2 e^{\beta \alpha^2 \sum_{j\in V(i)}(n_i-1)} \cosh(\beta \sum_{j\in V(i)} m_i)} \, ,
\label{SCE1local} \\
n_i &= \frac{2 e^{\beta \alpha^2 \sum_{j\in V(i)} (n_j-1)} \cosh(\beta \sum_{j\in V(i)} m_j)}{1+2 e^{\beta \alpha^2 \sum_{j\in V(i)}(n_i-1)} \cosh(\beta \sum_{j\in V(i)} m_i)} \, ,
\label{SCE2local}
\end{align}
\end{subequations}
which we will refer to as local SCEs.

\subsection{Uniform degree}
In the case in which all nodes have the same degree $k_i=z$, they will also have the same 
$m_i=m$ and $n_i=n$. Thus the free-energy function takes the simple form
\beqn
{\cal L}(m,n,\beta) &=& 
-\frac{z}{2} (m^2 + \alpha^2 n^2) + z \alpha^2 n \nonumber \\ &+&\frac{1}{\beta}
\left[ \frac{n+m}{2} \ln \frac{n+m}{2} + \frac{n-m}{2} \ln \frac{n-m}{2} + 
(1-n) \ln (1-n) \right]
\label{eq:landau}
\eeqn
and the (global) SCEs write (in the rest of this Appendix we redefine $\beta z$ as $\beta$ for brevity)
\begin{subequations}
\begin{align}
m &= \frac{2 e^{\beta \alpha^2 (n-1)} \sinh(\beta m)}{1+2 e^{\beta \alpha^2 (n-1)} \cosh(\beta m)} \, ,
\label{SCE1} \\
n &= \frac{2 e^{\beta \alpha^2 (n-1)} \cosh(\beta m)}{1+2 e^{\beta \alpha^2 (n-1)} \cosh(\beta m)} .
\label{SCE2}
\end{align}
\end{subequations}

If $m \neq 0$, the ratio between these two equations gives $n = m\coth(\beta m)$ and, substituting 
this into Eq.(\ref{SCE2}), we obtain a self-consistency equation for $m$ only:
\begin{equation}
m = \frac{2 e^{\beta \alpha^2 [m \coth(\beta m)-1]} \sinh(\beta m)}{1+2 e^{\beta \alpha^2 [m \coth(\beta m)-1]} \cosh(\beta m)}\,.
\label{eq:fully_eq}
\end{equation}
Expanding the latter around $m=0$ gives 
\beq
a_2 m + 2a_4 m^3 + 3a_6 m^4 +\dots  = 0
\label{eq:fully_coeff_1}
\eeq
where
\beq
a_2 = \frac{1}{2}\left( 1 - \frac{2\beta}{2 + e^{\alpha^2(\beta - 1)}} \right) , \quad
a_4 = \frac{\beta^3}{12} \frac{\left[4 - (1 + 2\alpha^2) e^{{\alpha^2(\beta - 1)}} \right]}{(2 + e^{\alpha^2(\beta - 1)})^2},
\label{eq:fully_coeff_2}
\eeq
and $a_6>0$. We thus have a tricritical point at the values $\alpha=\alpha_{tc}$, $\beta=\beta_{tc}$ that are solutions of the two equations $a_2 = a_4 = 0$ simultaneously, which gives
\begin{equation}
\ln [2(\beta_{tc} - 1)] = \frac{3 - \beta_{tc}}{2},
\label{eq:fully_tric}
\end{equation}
the numerical solution of which is $\beta_{tc} = 1.87676\dots$, and
\beq
\alpha_{tc} = \left(\frac{3 - \beta_{tc}}{2(\beta_{tc} - 1)}\right)^{1/2} = 0.800354 \dots\, .
\eeq

\subsection{Continuous transition}

For $\alpha < \alpha_{tc}$, the critical inverse temperature $\beta_c(\alpha)$ is determined by 
the condition $a_2=0$, $a_4 > 0$ which gives
\beq
\alpha^2 = \frac{1}{\beta_c(\alpha)-1} \ln \left[2 (\beta_c(\alpha) -1) \right]\,.
\eeq
The corresponding phase boundary is represented in Fig.~\ref{fully_phase} of the main text.
It is easy to check that $a_4>0$ in an interval of $\beta$ from zero to a value larger than
$\beta_c(\alpha)$. 

For $\beta$ near $\beta_c(\alpha)$ the spontaneous magnetization goes to zero with the usual mean-field exponent 1/2, i.e.
\beq
m = \left( \frac{-a_2}{2 a_4} \right)^{1/2} \underset{\beta \to \beta_c(\alpha)^+}\sim c(\alpha) \left( 
\frac{\beta}{\beta_c(\alpha)}-1 \right)^{1/2}
\label{eq:mcrit}
\eeq
where 
\beq
c(\alpha) =  \left( \frac{-6.0}{\beta_c(\alpha)}\frac{(\beta_c(\alpha)-1) \alpha^2-1}{2-(2 \alpha^2+1) [\beta_c(\alpha)-1]}\right)^{1/2} \,.
\eeq

At $\alpha=\alpha_{tc}$ the magnetization vanishes with the usual mean-field exponent 1/4:
\be
m = \left( -\frac{a_2}{3 a_6} \right)^{1/4} \underset{\beta\to \beta_{tc}^+}\sim c(\alpha_{tc}) \left( 
\frac{\beta}{\beta_c(\alpha)}-1 \right)^{1/4} \, 
\ee
with $c(\alpha_{tc}) = 1.03188$. 

The fraction of non-neutral agents, $n=1-n_0$, is given by $n = m \coth(\beta m)$ in the low temperature
phase, where $m$ is the solution of Eq.\eqref{eq:fully_eq}. In the high temperature phase, $n$ is found by solving numerically Eq.\eqref{SCE2} for $m=0$, i.e.
\be
n = \frac{2 e^{\beta \alpha^2 (n-1)}}{1+2 e^{\beta \alpha^2 (n-1)}} .
\label{SCEn}
\ee
The two solutions merge at $\beta_c(\alpha)$ and we have $\lim_{\beta\to \beta_c(\alpha)} n = \frac{1}{\beta_c(\alpha)}$ from both sides.

\subsection{Discontinuous transition}
For $\alpha > \alpha_{tc}$ there is a discontinuous transition at an inverse temperature $\beta_d(\alpha)$. 
The location of the transition is determined by finding the values of $m,n,\beta_d$ that satisfy 
simultaneously the two SCEs Eqs.\eqref{SCE1} and \eqref{SCE2} together with the condition 
that the free energy of the ferromagnetic and paramagnetic phases are equal, i.e.
${\cal L}(m, n, \beta_d) = {\cal L}(0, n(m=0,\beta_d), \beta_d)$,
where $n(m=0,\beta_d)$ is the solution of Eq.\eqref{SCEn}.  In practice we 
solved numerically Eq.\eqref{eq:fully_eq} starting from large $\beta$ and reduced $\beta$ in small
increments, following the solution until the above condition was met. This also allows us to
determine the limit of metastability of the ferromagnetic phase, namely the
value of $\beta$ below which there is no longer a non-zero solution of 
Eq.\eqref{eq:fully_eq}.

\section{Annealed mean-field approximation}\label{sec:AMF}
We now return to the case of non-uniform degrees $k_i$. In this case, in principle one could solve numerically the $2N$ local SCEs. Alternatively,
one can make the additional approximation consisting 
in treating the graph in an ``annealed'' fashion, by
 replacing the adjacency matrix $\epsilon_{i,j}$  (which is one if $i,j$ are connected and zero otherwise) by $k_i k_j/z N$, where $z=\langle k_i\rangle$ \cite{Bianconi_2002, Dorogovtsev_2008}.
If we introduce the weighted order parameters
\beq
m_w  =\frac{1}{z N}\sum_j k_j m_j,  \quad  n_w = \frac{1}{z N}\sum_j k_j n_j, 
\eeq
we have 
\beq
\sum_{j\in V(i)} m_j = \sum_{j\neq i} \epsilon_{i,j} m_j =
 \frac{1}{z N}\sum_{j\neq i} k_i k_j m_j = k_i m_w,
\eeq
and similarly $\sum_{j\in V(i)} n_j =  k_i n_w$. 
Multiplying the local SCEs by $k_i/(z N)$ and summing over $i$ we thus obtain two
SCEs for the weighted order parameters
\beqn
m_w &=& \frac{1}{z N}\sum_i k_i \frac{2 e^{\beta \alpha^2 k_i (n_w-1)} \sinh(\beta k_i m_w)}{1+2 e^{\beta \alpha^2 k_i (n_w-1)} \cosh(\beta k_i m_w)} \\
n_w &=& \frac{1}{z N}\sum_i k_i \frac{2 e^{\beta \alpha^2 k_i (n_w-1)} \cosh(\beta k_i m_w)}{1+2 e^{\beta \alpha^2 k_i (n_w-1)} \cosh(\beta k_i m_w)} 
\eeqn
For large $N$ we can replace the sum over $i$ with a sum over all possible degrees, 
\beqn
m_w &=& \frac{1}{z}\sum_{k} k P(k) \frac{2 e^{\beta \alpha^2 k (n_w-1)} \sinh(\beta k m_w)}{1+2 e^{\beta \alpha^2 k (n_w-1)} \cosh(\beta k m_w)} \\
n_w &=& \frac{1}{z}\sum_{k} k P(k) \frac{2 e^{\beta \alpha^2 k (n_w-1)} \cosh(\beta k m_w)}{1+2 e^{\beta \alpha^2 k (n_w-1)} \cosh(\beta k m_w)} .
\eeqn
Finally, after solving the above equations we can obtain the average order parameters as
\beqn
m &=& \frac{1}{N} \sum_i m_i = \sum_{k} P(k) \frac{2 e^{\beta \alpha^2 k (n_w-1)} \sinh(\beta k m_w)}{1+2 e^{\beta \alpha^2 k (n_w-1)} \cosh(\beta k m_w)} \\
n &=& \frac{1}{N} \sum_i n_i = \sum_{k} P(k) \frac{2 e^{\beta \alpha^2 k (n_w-1)} \cosh(\beta k m_w)}{1+2 e^{\beta \alpha^2 k (n_w-1)} \cosh(\beta k m_w)} .
\eeqn

\section{Exact solution for the one dimensional lattice}
\renewcommand{\theequation}{C-\arabic{equation}}
We solve the model exactly on a 1D lattice with periodic boundary conditions using the transfer-matrix method. By ordering the states as $1, 0, -1$, the transfer matrix
$\mathcal{T}_{\sigma_i,\sigma_{i+1}}=\exp\{\beta \left[\sigma_i\sigma_j+\alpha^2 (\sigma^2_i-1)(\sigma^2_j-1)\right]\}$ writes
\begin{equation}
\mathcal{T} = 
\begin{bmatrix}
w & 1 & w^{-1} \\
1  & v   & 1 \\
w^{-1} & 1 & w
\end{bmatrix},
\label{eq:1D_transfer}
\end{equation} 
where we defined $w \equiv e^{\beta}, v \equiv e^{\beta \alpha^2 }$. The free energy is given by 
$F(\beta, \alpha,N) = -\frac{1}{\beta }\ln {\mbox{tr}} \,\mathcal{T} = 
-\frac{1}{\beta} (\lambda_+^N + \lambda_0^N + \lambda_-^N)$, where the eigenvalues of $\mathcal{T}$ are
\begin{eqnarray}
\lambda_0 &=& w - w^{-1} \nonumber \\
\lambda_\pm &=&\frac{1}{2} \left[ w + w^{-1} + v\right] \pm \frac{1}{2}\left[ (w + w^{-1} - v)^{2} + 8\right]^{1/2} 
\label{eq:fully_system}
\end{eqnarray}
In the $N\to \infty$ limit the largest eigenvalue $\lambda_+$  dominates the sum, thus the free energy per spin is
\begin{equation}
f(\beta,\alpha) = \lim_{N\to\infty}\frac{1}{N}F(\beta,N)= -\frac{1}{\beta}\ln \left[ w + w^{-1} + v + \sqrt{(w + w^{-1} - v)^2 + 8} \right] \,.
\label{eq:1D_free_2}
\end{equation} 
Substituting $w$ and $v$ in the expression above, we obtain Eq.(\ref{eq:1Df}) of the main text.
The internal energy per spin is given by 
\begin{equation}
u(\beta, \alpha) = \frac{\partial (\beta f)}{\partial \beta} = - \frac{w - w^{-1} + \alpha^2 v + B^{-1/2}(w - w^{-1} - \alpha^2 v)(w + w^{-1} - v)}{w + w^{-1} + v +  B^{-1/2}}
\label{eq:1D_internal}
\end{equation} 
where $B \equiv (w + w^{-1} - v)^2 + 8$.
We also use the transfer matrix method to compute the spin correlation function
\beq
C(r) = \langle \sigma_i \sigma_{i+r}\rangle = \frac{{\mbox{tr}} (\Sigma \,\mathcal{T}^r \,\Sigma\, \mathcal{T}^{N-r})}{
{\mbox{tr}}\mathcal{T}^N}
\eeq
where $\Sigma$ is a diagonal matrix with elements $\Sigma_{\sigma,\sigma^\prime} = \sigma \delta_{\sigma,\sigma^\prime}$.
We can write the transfer matrix as $\mathcal{T} = \sum_\mu \lambda_\mu |\mu \rangle \langle \mu|$ where
$\mu=0,\pm 1$ and $|\mu\rangle$ are the eigenvectors of $\mathcal{T}$:
\beq
|0\rangle = \frac{1}{\sqrt{2}}\begin{bmatrix}
1\\ 0\\-1
\end{bmatrix}, \quad |\pm\rangle = \frac{1}{\sqrt{1+y_\pm^2}}\begin{bmatrix} 1\\ y_{\pm}\\1\end{bmatrix},
\eeq
where $y_\pm = \lambda_{\pm} - w - w^{-1}$.
In this way we obtain
\beq
C(r) = \frac{1}{\sum_\mu \lambda_\mu^N}
\left\{
\left[ \lambda_0^{N}  \left( \frac{\lambda_+}{\lambda_0}\right)^r 
+\lambda_+^{N}  \left( \frac{\lambda_0}{\lambda_+}\right)^r \right]
|\langle +|\Sigma|0\rangle|^2
+\left[ \lambda_0^{N}  \left( \frac{\lambda_-}{\lambda_0}\right)^r 
+\lambda_-^{N}  \left( \frac{\lambda_0}{\lambda_-}\right)^r \right] |\langle -|\Sigma|0\rangle|^2 \right\}
\eeq
which, since $|\lambda_+|>|\lambda_0|>|\lambda_-|$, for $N\to \infty$ gives
\beq
C(r)  = \frac{4}{2+y_+^2} e^{-r /\xi}, \quad  \xi \equiv \frac{1}{\ln \frac{\lambda_+}{\lambda_0}},
\eeq
where we used $|\langle +|\Sigma|0\rangle|^2 = 4/(2+y_+^2)$.
Inserting Eqs.(\ref{eq:fully_system}) into the expression above for $\xi$, 
and substituting $w$ and $v$, we obtain Eq.(\ref{xi}) of the main text.

\section{Residual Energy for the Erd\"os-R\'enyi graphs}\label{sec:AppER}
\renewcommand{\theequation}{S-\arabic{equation}}
\renewcommand\thefigure{\thesection.\arabic{figure}} 
\setcounter{figure}{0}

\begin{figure}[h!]
\centering
\includegraphics[width=\columnwidth]{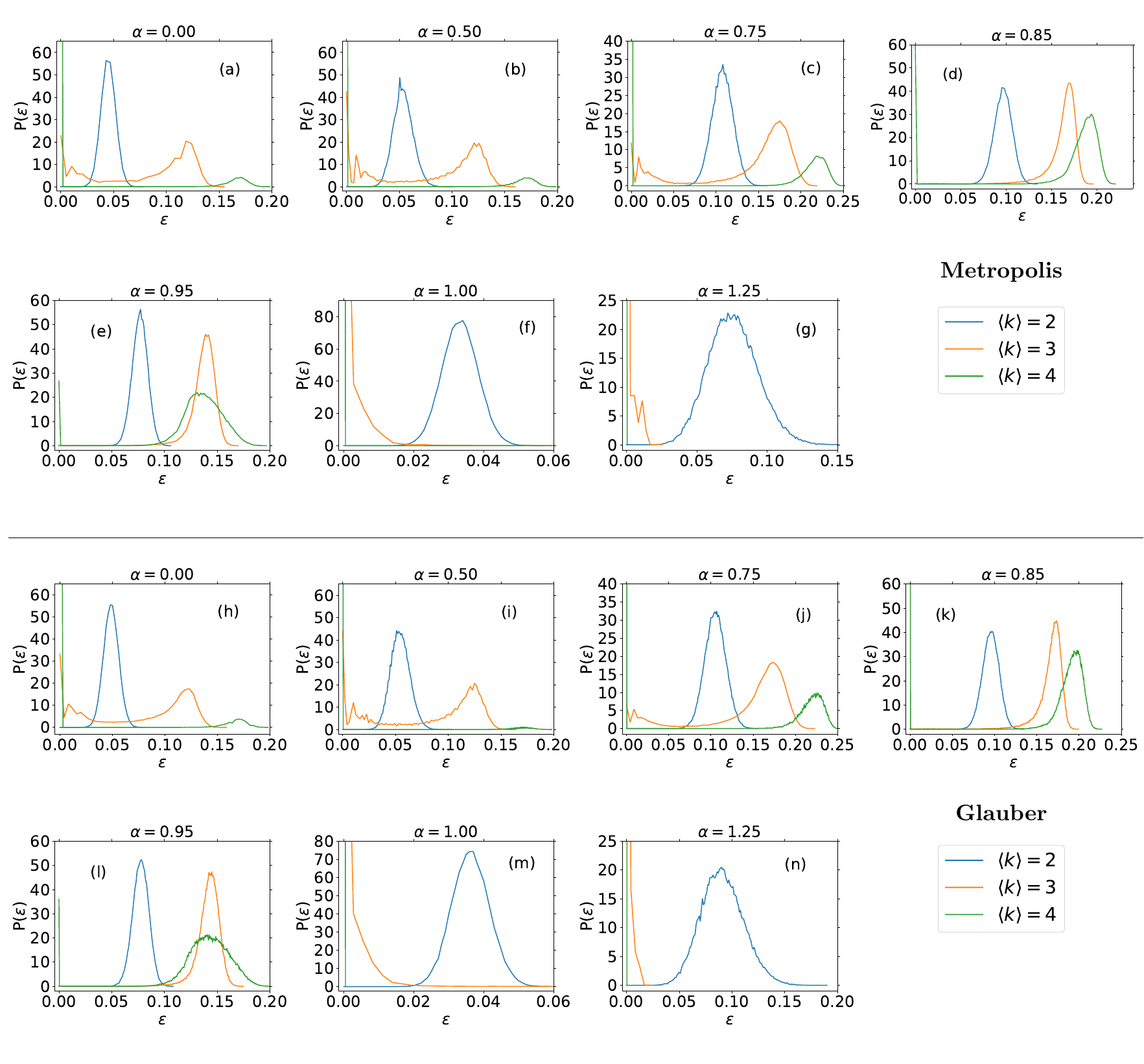}
\caption{\small Probability distribution of the residual energy for different values of $\alpha$ and $\langle k \rangle = 2, 3, 4$, using  Metropolis 
(top panel) and  Glauber dynamics (bottom panel). Results are obtained for a single realization of a network with $N=500$, and $10^5$ repetitions of the simulation.}
\label{ER_res_500}
\end{figure}

In this section we present additional zero-temperature MC results for ER graphs with low connectivity, in order to explore the effects of dynamical traps that prevent the system from reaching optimality. We considered both the Metropolis and Glauber dyamics, and considering that the system has reached a steady state when the energy does not change for $100\log(N)$ MCS.
with the same energy, then calculating the difference between the measured energy and the ground state. 
Fig.~\ref{ER_res_500} shows the distribution of the residual energy $\epsilon = (E_{steady} - E_{0})/N z$. 
We can observe that both algorithms give almost identical results.

If we compare the case $\alpha = 0$ with the Ising model \cite{Baek2012} we appreciate that the peaks are lower in our model, due to the existence of the neutral opinion, which reduces the amount of energy necessary to transition between states $+1$ and $-1$. For the case $\langle k \rangle = 4$, which presents a behavior very similar to the complete graph, deviations from the ground state are only relevant when we are close to the mean-field tricritical point, pointing to the existence of a first order transition also in ER graphs.  

\begin{figure}[h!]
\centering
\includegraphics[width=\columnwidth]{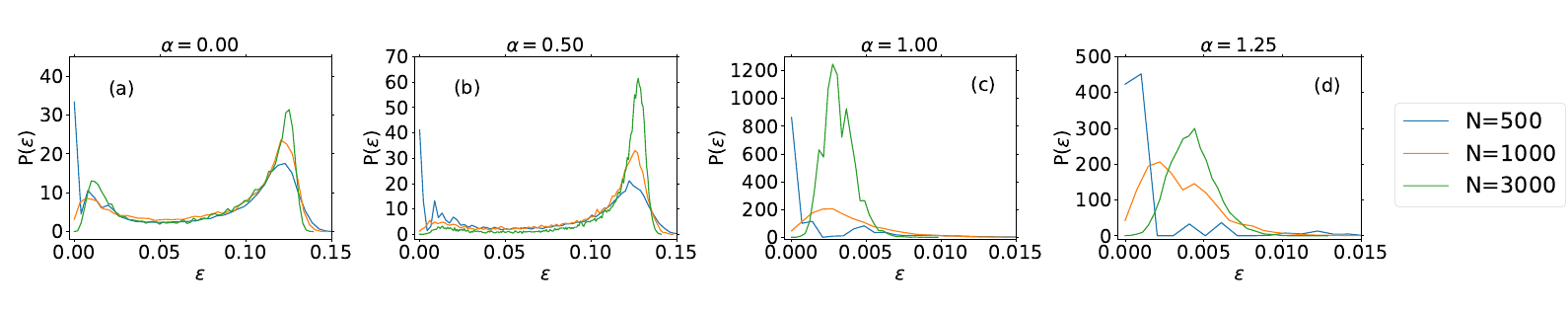}
\caption{\small{Probability distribution of the residual energy for ER graphs with $\langle k \rangle = 3$ and sizes $N=500$ and $N=1000$.}}
\label{ER_res_Ncompare}
\end{figure}

%\vspace{10cm}
%
In Fig.~\ref{ER_res_Ncompare} we can compare the probability distribution of the residual energy for two ER networks of different sizes and $\langle k \rangle = 3$ using Glauber dynamics. The distribution is bimodal for values of $\alpha \leq 0.75$, but becomes unimodal for larger values of $\alpha$, which is related to the symmetry breaking caused by the neutral opinion above this threshold. 
\end{appendix}
\end{document}